\documentclass[a4paper,11pt, twocolumn]{quantumarticle}
\pdfoutput=1

\usepackage{amsthm,amsmath,amssymb,graphicx,comment} 
\usepackage{bbold,array}
\usepackage[none]{hyphenat}
\usepackage{dsfont}

\usepackage{microtype,amsbsy}
\usepackage{units}
\usepackage{multirow}

\usepackage{cite}
 
\usepackage[usenames,dvipsnames,table]{xcolor}
\definecolor{purple}{RGB}{128,0,128}
\definecolor{ultramarine}{RGB}{63, 0, 255}
\definecolor{medblue}{RGB}{0, 0, 100}
\definecolor{panblue}{RGB}{0,24,150}
\definecolor{carmine}{RGB}{150, 0, 24}
\definecolor{gray}{RGB}{150, 150, 150}
\definecolor{darkred}{RGB}{200, 0, 0}
\definecolor{darkgreen}{RGB}{0, 80, 0}
\definecolor{darkblue}{RGB}{0, 0, 200}

\definecolor{nred}{rgb}{0.9,0.1,0.1}
\definecolor{nblack}{rgb}{0,0,0}
\definecolor{nblue}{rgb}{0.2,0.2,0.8}
\definecolor{ngreen}{rgb}{0.2,0.6,0.2}

\newcommand{\bomega}{\boldsymbol{\omega}}
\newcommand{\bs}[1]{\boldsymbol{#1}}

\newcommand{\GPTt}{\ensuremath{\Omega}}
\newcommand{\gpEff}{\ensuremath{\mathcal{E}}}
\newcommand{\gp}[1]{\ensuremath{\mathbf{#1}}}

\newcommand{\ketbra}[2]{\ensuremath{{\textstyle{\ket{#1}}\!{\bra{#2}}}}}

\usepackage{hyperref}
\hypersetup{colorlinks, breaklinks, linkcolor=quantumviolet, urlcolor=darkblue, citecolor=darkblue}

\usepackage{bm}

\newcommand{\ket}[1]{\left| #1 \right\rangle}
\newcommand{\bra}[1]{\left\langle #1 \right|}

\newcommand{\beq}{\begin{equation}}
\newcommand{\eeq}{\end{equation}}
\newcommand{\bea}{\begin{align}}
\newcommand{\eea}{\end{align}}
\newcommand*{\vertbar}{\rule[-1ex]{0.5pt}{2.5ex}}

\newtheoremstyle{formltn}% name
{}%\abovedisplayskip}% Space above
{}%\belowdisplayskip}% Space below
{\itshape}% Body font
{}% Indent amount
{\bfseries}% Theorem head font
{ of necessary conditions for existence of an ontological model}% Punctuation after theorem head
{\parindent}% Space after theorem head
{\thmname{#1} \thmnumber{{#2}}\thmnote{\normalfont#3}}% Theorem head spec (can be left empty, meaning ‘normal’)
\theoremstyle{formltn}
\newtheorem{formulation}{Formulation}
\newtheoremstyle{thepolytope}% name
{}%\abovedisplayskip}% Space above
{}%\belowdisplayskip}% Space below
{\itshape}% Body font
{}% Indent amount
{\bfseries}% Theorem head font
{ of the noncontextual measurement-assignment polytope:}% Punctuation after theorem head
{\parindent}% Space after theorem head
{\thmname{#1} \thmnumber{{#2}}\thmnote{\normalfont#3}}% Theorem head spec (can be left empty, meaning ‘normal’)
\theoremstyle{thepolytope}

\usepackage{etoolbox}
\newcounter{equationstore}
\AtBeginEnvironment{formulation}{
\setcounter{equationstore}{\value{equation}}
\setcounter{equation}{0}
}
\AtEndEnvironment{formulation}{
\setcounter{equation}{\value{equationstore}}}
\AtBeginEnvironment{thepolytope}{
\setcounter{equationstore}{\value{equation}}
\setcounter{equation}{0}
}
\AtEndEnvironment{thepolytope}{
\setcounter{equation}{\value{equationstore}}}

\usepackage{soul}
\usepackage{stmaryrd}

\begin{document}
\title{Noncontextuality inequalities for prepare-transform-measure scenarios}
\author{David Schmid}
\email{davidschmid10@gmail.com} 
\affiliation{International Centre for Theory of Quantum Technologies, University of Gda\'nsk, 80-309 Gda\'nsk, Poland}
\affiliation{Perimeter Institute for Theoretical Physics, Waterloo, Ontario, Canada, N2L 2Y5}
\author{Roberto D. Baldij\~ao}
\affiliation{International Centre for Theory of Quantum Technologies, University of Gda\'nsk, 80-309 Gda\'nsk, Poland}
\affiliation{Perimeter Institute for Theoretical Physics, Waterloo, Ontario, Canada, N2L 2Y5}
\author{John H.~Selby}
\affiliation{International Centre for Theory of Quantum Technologies, University of Gda\'nsk, 80-309 Gda\'nsk, Poland}
\author{Ana Bel\'{e}n Sainz}
\affiliation{International Centre for Theory of Quantum Technologies, University of Gda\'nsk, 80-309 Gda\'nsk, Poland}
\affiliation{Basic Research Community for Physics e.V., Germany}
\author{Robert W. Spekkens}
\affiliation{Perimeter Institute for Theoretical Physics, Waterloo, Ontario, Canada, N2L 2Y5}
\begin{abstract} 
We provide the first systematic technique for deriving witnesses of contextuality in prepare-transform-measure scenarios. More specifically, we show how linear quantifier elimination can be used to compute a polytope of correlations consistent with generalized noncontextuality in such scenarios. This polytope is specified as a set of noncontextuality inequalities that are necessary and sufficient conditions for observed data in the scenario to admit of a classical explanation relative to any linear operational identities, if one ignores some constraints from diagram preservation. While including these latter constraints generally leads to tighter inequalities, it seems that nonlinear quantifier elimination would be required to systematically include them. We also provide a linear program which can certify the nonclassicality of a set of numerical data arising in a prepare-transform-measure experiment.
We apply our results to get a robust noncontextuality inequality for transformations that can be violated within the stabilizer subtheory. Finally, we give a simple algorithm for computing all the linear operational identities holding among a given set of states, of transformations, or of measurements.
\end{abstract}

\maketitle

\newpage
\tableofcontents

\newpage
\section{Introduction} 

A well-motivated and universally applicable method for demonstrating that a theory or data set cannot be explained  classically is to prove that it cannot be reproduced in any generalized-noncontextual ontological model~\cite{gencontext}. This notion of classicality is motivated by a form of Leibniz's principle of the identity of indiscernibles~\cite{Leibniz}. It also coincides with the natural notion of classical explainability used in quantum optics, namely positivity of some quasiprobability representation~\cite{negativity,SchmidGPT,Schmid2024structuretheorem}, and {\em also} coincides with the natural notion of classical explainability in the framework of generalized probabilistic theories~\cite{SchmidGPT,ShahandehGPT,Schmid2024structuretheorem}, namely linear embeddability into a simplicial theory. In the latter framework, the simplicity and naturalness of this notion of classical explainability is made especially manifest: {\em Given a circuit of gates on any set of systems, such a classical explanation associates to each system a random variable and (linearly) associates to each gate a stochastic map over the random variables associated with the gate's inputs, such that the composition of the stochastic maps reproduces the predictions of the circuit}~\cite{Schmid2024structuretheorem}.

A major advantage of this notion of nonclassicality over other standard notions is that it has {\em universal applicability}. Given {\em any} experiment or theory, one can always ask whether or not there exists any noncontextual model for it. This is not true of violations of Bell inequalities~\cite{Bell} (or more general causal compatibility inequalities~\cite{inflation}), which apply only to specific causal structures. Nor is it true of Kochen-Specker noncontextuality~\cite{KS}, which refers only to the structure of measurements, not transformations or states (much less full theories). Macroscopic realism as traditionally defined~\cite{AJLeggett_2002} also applies only in particular scenarios (e.g., those with sequences of measurements). However, it has been shown that a simpler yet more sophisticated definition of macroscopic realism can be given in the framework of generalized probabilistic theories, and this notion can be applied universally in the same way that generalized noncontextuality can~\cite{Schmid2024reviewreformulation}. Still, macroscopic realism is (obviously) not well motivated for systems that are not macroscopic, so in practice it is very rarely relevant in actual quantum experiments (and even when it is, generalized noncontextuality is equally relevant and very closely linked to it~\cite{Schmid2024reviewreformulation}).

The universal applicability of this notion of nonclassicality allows for the unified study, comparison, and classification of arbitrary phenomena within quantum theory---or indeed {\em any} operational theory---in terms of which are classical and which are nonclassical.  Such studies 
have been carried out for phenomena relating to computation~\cite{Schmid2022Stabilizer,shahandeh2021quantum}, state discrimination~\cite{schmid2018contextual,flatt2021contextual,mukherjee2021discriminating,Shin2021}, interference~\cite{Catani2023whyinterference,catani2022reply,catani2023aspects}, compatibility~\cite{selby2023incompatibility,selby2023accessible,PhysRevResearch.2.013011}, uncertainty relations~\cite{catani2022nonclassical}, metrology~\cite{contextmetrology}, thermodynamics~\cite{contextmetrology,comar2024contextuality}, weak values~\cite{AWV, KLP19}, coherence~\cite{rossi2023contextuality,Wagner2024coherence}, quantum Darwinism~\cite{baldijao2021noncontextuality}, information processing and communication~\cite{POM,RAC,RAC2,Saha_2019,Yadavalli2020,PhysRevLett.119.220402,fonseca2024robustness}, cloning~\cite{cloningcontext}, broadcasting~\cite{jokinen2024nobroadcasting}, and (as mentioned above) Bell~\cite{Wright2023invertible,schmid2020unscrambling,zhangent} and Kochen-Specker scenarios~\cite{operationalks,kunjwal2018from,Kunjwal16,Kunjwal19,Kunjwal20,specker,Gonda2018almostquantum}.

Among proofs that the operational statistics of some experimental set-up fail to admit  of a noncontextual\footnote{Here and henceforth, we use the shorthand term {\em noncontextual} to mean generalized noncontextual.} ontological model, an important distinction is whether they are noise-robust or not.

Proofs that have the form of a no-go theorem generally only demonstrate that certain {\em idealized} operational statistics predicted by quantum theory cannot be realized in a noncontextual ontological model.
But such proofs generally do not apply  
to noisy experimental data because such data  may deviate significantly from the idealized predictions.
A common route to {\em noise-robust} tests of the existence of a noncontextual ontological model are {\em noncontextuality inequalities}~\cite{POM,kunjwal2018from,mazurek2016experimental,robust}.  
These are constraints on the operational statistics that are satisfied if and only if a noncontextual ontological model of the statistics exists. (Another route is via searching for simplex-embeddings~\cite{SchmidGPT,selby2024linear}.)
Such tests also provide a means of defining the classical-nonclassical boundary for the statistics in an arbitrary operational theory, including alternatives to quantum theory~\cite{Hardy,GPT_Barrett}.

A second important distinction among proofs concerns the nature of the experimental set-up: 
Most existing proofs posit a set-up wherein a system is subjected to one of a set of preparation procedures followed by one of a set of measurement procedures, a form that is termed a {\em prepare-measure scenario}.  
There are some proofs that consider a preparation stage followed by a sequence of measurements, where at each step of the sequence, one of a set of measurements is implemented)~\cite{lostaglio2018,KLP19,Anwer2021noiserobust,comar2024contextuality,AWV,PP2}.  
There are also some proofs wherein one of a set of transformation procedures is implemented between the preparation and the measurement stages of the experiment, a form that is termed a {\em prepare-transform-measure scenario} and which is depicted in Fig.~\ref{Fig:OntologicalModelPTMScenario}(a)~\cite{gencontext,negativity,Lillystone2019,Schmid2022Stabilizer,cloningcontext,Wagner2024coherence}.\footnote{There is even some recent work on proofs in experimental set-ups with a generic circuit structure~\cite{Schmid2024structuretheorem,Schmid2022Stabilizer}.}

This work is concerned with deriving noise-robust noncontextuality inequalities for this last type, namely, prepare-transform-measure scenarios. We follow an approach like that of Ref.~\cite{Schmid2018}, which gave an algorithm for determining the full set of noncontextuality inequalities for a prepare-measure scenario relative to a fixed set of linear operational identities among the states and a fixed set of linear operational identities among the effects.  
The present manuscript aims to extend these ideas to prepare-transform-measure scenarios, by providing an algorithm for deriving noncontextuality inequalities in any such scenario, relative to any fixed set of linear operational identities. These inequalities are necessary but not sufficient for the existence of a noncontextual model relative to these operational identities, since our algorithm does not take into account all possible constraints arising from compositionality (such as from sequences of transformations or consideration of multiple copies of a system), as we discuss in Section~\ref{sec:opidentities} and immediately after the linear program (Formulation~\ref{c2}).

In addition, we give a linear program that directly certifies whether any given set of experimental data from a prepare-transform-measure scenario admits of a noncontextual ontological model. If it does admit of one, the program provides a candidate classical model for the scenario; otherwise, it provides a noncontextuality inequality that is maximally violated in the scenario. This circumvents the need to derive an entire polytope (set of inequalities) for the given scenario, and so is more computationally efficient.

We then give an illustrative example of the application of these tools. 
 Specifically, we find the first
 noise-robust noncontextuality inequality for transformations that can be violated within the stabilizer theory~\cite{gottesman1997stabilizer,gottesman1998heisenberg} for a single qubit.

Another contribution of this work is to provide systematic techniques for deriving all possible operational identities that have a linear form. This is an important supplement to the algorithm in this work and also to prior works (most notably  Ref.~\cite{Schmid2018}). In Appendix~\ref{appendix:AllOpIdentitiesTransf}, we show how one can identify a finite set of  linear  operational identities among states which imply {\em all} linear operational identities among the states---that is, a {\em generating set} of such identities. We then show how to identify a generating set of linear operational identities for the effects and for the transformations as well. 

Combining these arguments with the algorithm in this paper---or with that in  Ref.~\cite{Schmid2018}---implies that one need not specify any specific operational identities among states or among effects or among transformations. Rather, one can now merely stipulate the set of states, effects, and transformations in one's scenario, and then use the technique described in Appendix~\ref{appendix:AllOpIdentitiesTransf} to find a generating set of linear operational identities that hold among the elements of each set, and use these as the input to the program. This gives stronger noncontextuality inequalities than one would obtain by using a strict subset of the operational equivalences.

A final conceptual contribution of this work is in Section~\ref{sec:opidentities}, where we discuss how not all operational identities are of a linear form. This is most obviously true for transformations, where one has operational identities that arise from sequential composition. However, it is also true for states and effects, although this fact has largely been overlooked in past works (with the exception of Ref.~\cite{schmid2024addressing}). We give some discussion and examples to illustrate this fact.
Neither this work nor prior work provide methods for dealing with operational identities that are not of a linear form. While it would be desirable to also include all possible constraints of this sort,  any technique that does so will likely be computationally very demanding.

In total, this work constitutes the first systematic technique for testing nonclassicality in a noise-robust manner outside of prepare-measure scenarios, and so constitutes a first step towards the long-term goal of understanding nonclassicality as a resource for quantum computation.

\section{Generalized probabilistic theories and ontological models thereof}

We first briefly review some background concepts, following the presentation in Ref.~\cite{SchmidGPT}. 

\subsection{Generalized probabilistic theories}

\begin{figure}[htb!]
\includegraphics[width=0.35\textwidth]{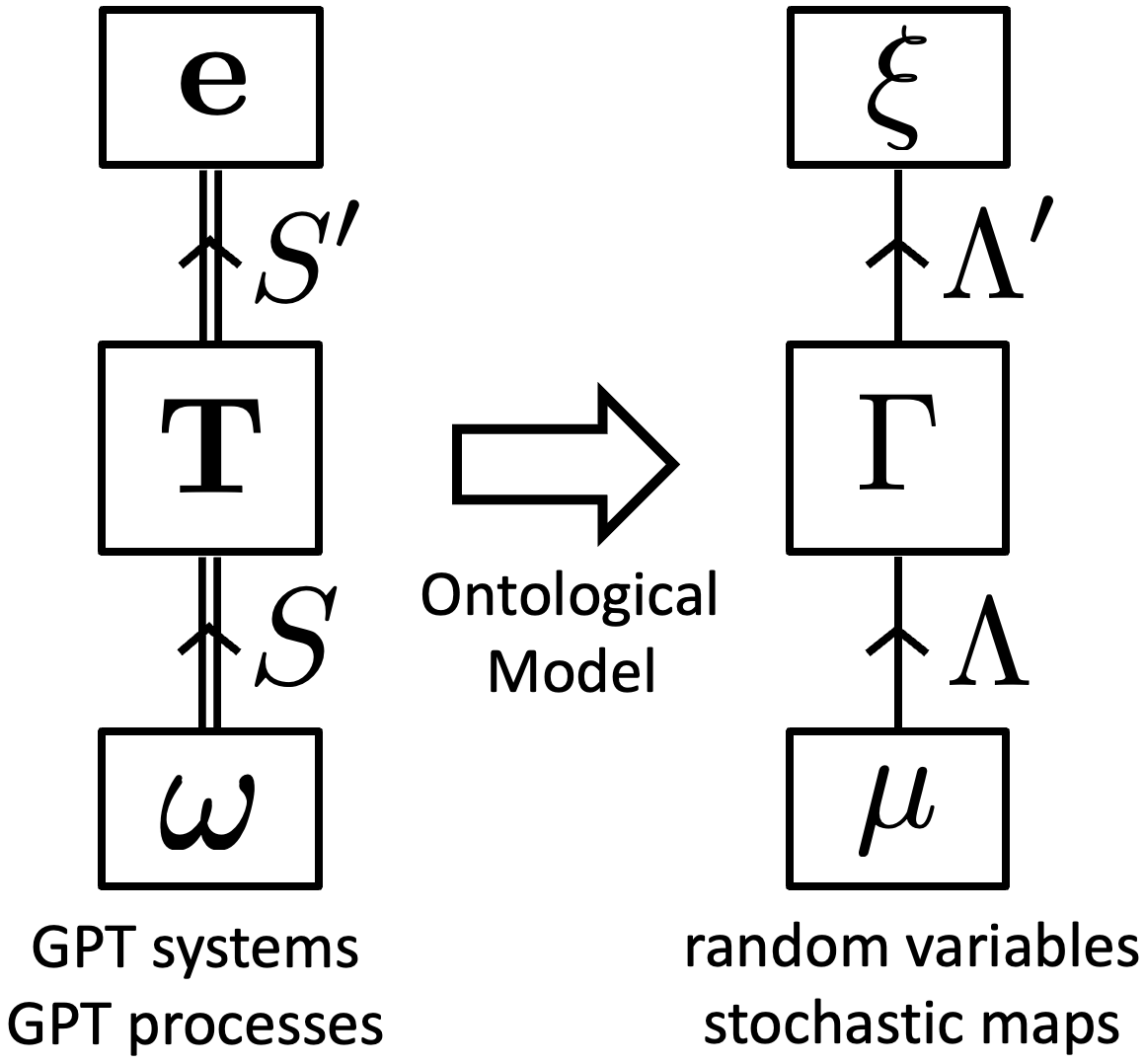}
\caption{(a) The scenario of interest in this paper. (b) An ontological model for the scenario in (a).}
\label{Fig:OntologicalModelPTMScenario}
\end{figure}

The framework of generalized probabilistic theories (GPTs) provides a means of describing the landscape of possible theories of the world, as characterized (solely) by  the operational statistics they predict~\cite{Hardy,GPT_Barrett}.  
 Quantum and classical theories are included as special cases, but the framework also accommodates alternatives to these.
Although the framework allows for arbitrary sequential and parallel composition of processes, we will focus on the fragment of a GPT representing the composition of a single state, transformation, and measurement, as in Figure~\ref{Fig:OntologicalModelPTMScenario}(a). We refer to this as a $\mathcal{PTM}$ scenario.

A given GPT associates to a system a convex set of {\em states}, denoted $\mathcal{P}$. 
One can think of this set as being a generalization of the Bloch ball in quantum theory, where the states in the set are the normalised (potentially mixed) states of the theory. We make the standard assumptions that $\mathcal{P}$ is finite dimensional and compact. While $\mathcal{P}$ naturally lives inside an affine space, $\mathsf{AffSpan}[\mathcal{P}]$, for convenience we will represent it as living inside a real vector space  $V$ of one dimension higher, where we embed $\mathsf{AffSpan}[\mathcal{P}]$ as a hyperplane in $V$ which does not intersect with the origin $\boldsymbol{0}$. This is analogous to embedding the Bloch-Ball within the real vector space of Hermitian matrices. 
 
 Note that we will {\em not} restrict attention to GPTs satisfying the no-restriction hypothesis~\cite{chiribella2010probabilistic}, which stipulates that all the states and effects that are logically possible must also be physically possible.

A GPT also associates to every system a set of GPT {\em  effects, }
$\mathcal{E}$,  which live in the dual vector space $V^*$.  In the framework of GPTs, the probability of obtaining an effect $\boldsymbol{e}\in\mathcal{E}$
given a state $\bomega\in\mathcal{P}$ is given by:
\beq
p(\boldsymbol{e},\bomega) :=  \boldsymbol{e}\circ \bomega = \boldsymbol{e}(\bomega). 
    \eeq
 We require that $\mathcal{E}$ must satisfy the following constraints.
If one defines the dual of $\mathcal{P}$, denoted $\mathcal{P}^*$,  as the set of vectors in  $V^*$ whose %inner product with 
evaluation on  all state vectors in $\GPTt$ is between $0$ and $1$, i.e.,
\beq
\mathcal{P}^* := \{ \boldsymbol{x}\in V^* |
\boldsymbol{x}\circ \bomega \in [0,1] \ \forall \bomega \in \mathcal{P}\},
\eeq
\sloppy then $\mathcal{E}$ is a compact convex set contained in $\mathcal{P}^*$, ${\mathcal{E} \subseteq \mathcal{P}^*}$,
which  contains the origin $\boldsymbol{0}$ and the ``unit effect'' $\boldsymbol{u}$, which in turn satisfy, respectively,  
$\boldsymbol{0} \circ \bomega = 0$ and $\boldsymbol{u}\circ\bomega = 1$ for all  $\bomega\in\mathcal{P}$, and, that for every $\boldsymbol{e}\in \mathcal{E}$ there exists $\boldsymbol{e}^\perp\in \mathcal{E}$ such that $\boldsymbol{e}+\boldsymbol{e}^\perp = \boldsymbol{u}$. 
Due to how we embedded $\mathsf{AffSpan}[\mathcal{P}]$ within $V$, $\boldsymbol{u}$ necessarily exists and is unique~\cite{chiribella2010probabilistic}. 
A {\em measurement} in a GPT is a set of effects that sums to the unit effect.  Note that, due to the demanded existence of $\boldsymbol{e}^\perp$, every effect necessarily belongs to some measurement, namely $\{\boldsymbol{e},\boldsymbol{e}^\perp\}$. We will follow the common practice of treating measurements as sets of effects (rather than as channels from a GPT system to a classical outcome system).

The state and effect spaces of any valid GPT must satisfy the principle of {\em tomography}, which states that the GPT states and GPT effects can be uniquely identified by the probabilities that they produce. Formally, for the GPT states, we have  that  
$\gp{e}\circ\gp{s_1} = \gp{e}\circ\gp{s_2}$
for all $\gp{e}\in\gpEff$ if and only if $\gp{s_1} = \gp{s_2}$, and for the GPT effects, we have that
$\gp{e_1}\circ\gp{s} = \gp{e_2}\circ\gp{s}$ 
for all $\gp{s}\in\mathcal{P}$ if and only if $\gp{e_1}  = \gp{e_2}$.

A GPT transformation is a linear map\footnote{Strictly speaking, this definition assumes tomographic locality (without which a single transformation would be defined as a family of linear maps). We make this assumption for simplicity, but our results do not rely on it (although the particular operational identities among one's transformations will depend on it).} which takes GPT states to GPT states and (by precomposition) GPT effects to GPT effects. The probability of obtaining an effect $\boldsymbol{e}\in\mathcal{E}$ given a state $\bomega\in\mathcal{P}$  and transformation $\boldsymbol{T}\in \mathcal{T}$ is given by 
\beq
p(\boldsymbol{e},{\bf T},\bomega)
:=  \boldsymbol{e}\circ {\bf T} \circ \bomega.
\eeq
Similarly to states and effects, these satisfy the principle of {\em tomography}\footnote{ Again, strictly speaking what we are defining here is local tomography for simplicity of presentation, to formally define things without this assumption would require introducing all of the formalism of composite systems for GPTs.}, that is, they can be uniquely identified by the probabilities that they produce: $\boldsymbol{e}\circ {\bf T_1} \circ \bomega =\boldsymbol{e}\circ {\bf T_2} \circ \bomega$ for all $\bomega \in \mathcal{P}$ and $\boldsymbol{e}\in \mathcal{E}$ if and only if ${\bf T_1} = {\bf T_2}$. Within $\mathcal{T}$ we can define discard-preserving transformations, ${\bf N}$ as those that satisfy $\boldsymbol{u}\circ {\bf N} = \boldsymbol{u}$ and it must be the case that for every transformation ${\bf T}\in \mathcal{T}$ there exists  a transformation ${\bf T}^\perp\in \mathcal{T}$ such that ${\bf T}+{\bf T}^\perp$ is discard-preserving.

\subsection{Ontological models of GPTs}

An ontological model of a GPT is an attempt to provide a causal explanation of   the operational statistics in terms of a space $\Lambda$ of ontic states  for the system. Very simply, such an explanation takes the form of a linear and diagram-preserving~\cite{Schmid2024structuretheorem} map that associates to each GPT system a classical random variable ($\Lambda$), and then represents each GPT process (e.g., a state, transformation, or measurement) as a substochastic map from the random variables associated with its inputs to those associated with its outputs. Such a mapping is shown for a $\mathcal{PTM}$ scenario in Figure~\ref{Fig:OntologicalModelPTMScenario}. 

The precise definition of an ontological model for a fully compositional GPT was first given in Ref.~\cite{Schmid2024structuretheorem}. Here, we will give a simplified definition for the special case of $\mathcal{PTM}$ scenarios: an ontological model of a GPT associates to each GPT state $\bomega \in \mathcal{P}$ a probability distribution (often called an {\em epistemic state}) over $\Lambda$, denoted $\mu_{\bomega}$, to each GPT effect $\boldsymbol{e} \in \mathcal{E}$ a response function on $\Lambda$, denoted $\xi_{\boldsymbol{e}}$, and to each GPT transformation ${\bf T}$ a stochastic map $\Gamma_{\bf T}(\lambda'|\lambda)$. Moreover, these representations must jointly reproduce the probability rule of the GPT by satisfying 
\begin{equation}\label{OMGPTprob}
\boldsymbol{e}\circ{\bf T}\circ\bomega  = \int_{\Lambda,\Lambda'} d\lambda d\lambda' \xi_{\boldsymbol{e}}(\lambda') \Gamma_{\bf T}(\lambda'|\lambda)\mu_{\bomega}(\lambda).
\end{equation}
Finally, an ontological model must represent the identity transformation by the ontic identity $\delta_{\lambda',\lambda}$, and the unit effect by the all-ones vector (the unit effect in any ontological theory).

\subsection{The connection to noncontextuality} \label{NCandGPTs}

Throughout this paper, we assume one has a GPT representation of the states, transformations, and effects in one's scenario. This is in contrast to many earlier works on generalized noncontextuality, which instead take unquotiented operational theories (with laboratory preparation procedures, measurement procedures, and so on) as their starting point. Either starting point is equally meaningful, but quotiented theories have a host of technical advantages in practice. Most notably, processes in a quotiented theory naturally live in a vector space, while laboratory procedures do not. Moreover, beginning with the quotiented theory means we do not need to consider sets of processes that are tomographically complete, as would otherwise be required~\cite{schmid2024addressing,PuseydelRio}. (Although, of course,  exactly this same assumption of tomographic completeness is required to give a GPT representation of a given set of processes~\cite{chiribella2010probabilistic,Schmid2024structuretheorem}, so this is a practical advantage, not a conceptual one.)

As discussed in detail in Ref.~\cite{Schmid2024structuretheorem}, an ontological model of a GPT cannot be said to be contextual or noncontextual. However, it is also shown there that an ontological model of a GPT exists if and only if a noncontextual ontological model exists for the unquotiented operational theory from which the GPT arose. It is through this connection that we are able to focus entirely on ontological models of GPTs, while nonetheless drawing conclusions about noncontextuality.  This bridge will be used (usually implicitly) throughout the work.

Note that one can {\em always} construct an ontological model of an unquotiented operational theory by allowing this model to be generalized-contextual (using the analogue of the construction of   Ref.~\cite{Beltrametti_1995}).  But it is {\em not} the case that one can always construct an ontological model of a GPT, because such models do not have the benefit of the representational flexibility afforded by nontrivial context-dependences. Indeed, it is this lack of flexibility that implies the necessity of negativity in quasiprobability representations of some GPTs~\cite{negativity,Ferrie_2008}.

\subsection{Operational identities and constraints from linearity}\label{sec:opidentities}

The fact that an ontological model of a GPT is a {\em linear} map implies that it must preserve all linear equalities that hold among GPT processes. The fact that it is diagram-preserving implies that it must preserve all compositional equalities among GPT processes.  These linear and compositional equalities encode the convex geometry and compositional structure of the GPT, which in turn encodes all of its empirical content. Thus, the constraints from linearity are central to the question of whether or not an ontological model exists for a given GPT. 

We refer to every equality in a given GPT as an {\em operational identity} of that GPT. (This term mirrors the term {\em operational equivalence} that appears in the literature on noncontextuality---a term which applies to unquotiented operational theories.) 

Let us focus first on operational identities which are linear equalities among states. As a simple example when the GPT in question is quantum, the geometry of the Bloch ball implies that 
\begin{equation}
\frac12 \ket{0}\!\!\bra{0}+\frac12 \ket{1}\!\!\bra{1} = \frac12 \ket{+}\!\!\bra{+}+\frac12 \ket{-}\!\!\bra{-}.
\end{equation} 
In any ontological model of a qubit, then, the epistemic states associated with these four states must satisfy
\begin{equation}
\frac12 \mu_{\ket{0}\!\bra{0}}+\frac12 \mu_{\ket{1}\!\bra{1}} = \frac12\mu_{\ket{+}\!\bra{+}}+\frac12 \mu_{\ket{-}\!\bra{-}}.
\end{equation} 

Each linear equality among GPT processes provides a constraint which must be satisfied by any ontological model. 

Given a set of GPT states, the set of all linear equalities among them can be indexed by $a$ and written as
\begin{equation}
\left\{\sum_{\bomega \in \mathcal{P}} \alpha_{\bomega}^{(a)} \bomega =0\right\}_a,
\label{OpIdentitiesStates}
\end{equation}
where for each $a$ one has some set of real numbers $\{\alpha_{\bomega}^{(a)}\}_{\bomega}$. 
These coefficients can be found explicitly via simple linear algebra, as we show in Appendix~\ref{appendix:AllOpIdentitiesTransf}. (In general, there are an infinite number of these equalities, but one can nonetheless characterize all of these in an efficient way. In particular, there will always be a finite set of linearly independent operational identities, as we show in Appendix ~\ref{appendix:AllOpIdentitiesTransf}.) 
By linearity, any ontological model of the GPT (should one exist) must satisfy
\begin{equation}
\forall \lambda, a: \quad \sum_{\bomega\mathcal{P}} \alpha_{\bomega}^{(a)} \mu_{\bomega}(\lambda)=0.
\label{OMIdentitiesStates}
\end{equation} 

Similarly, one can write the set of all linear equalities holding among effects as 
\begin{equation}
\left\{\sum_{\bf e \in \mathcal{E}} \alpha_{\bf e}^{(b)} {\bf e} = 0\right\}_b,
\end{equation}
and the constraints on the ontological model from linearity are
\begin{equation}
\forall \lambda, b: \quad  \sum_{\bf e \in \mathcal{E}} \alpha_{\bf e}^{(b)} \xi_{\bf e}(\lambda) = 0.
    \end{equation}

Note that for effects (unlike for states) there are certain operational identities that are unavoidable; for instance, given any effect $\boldsymbol{e}$ in a scenario, one necessarily also has the effect $\boldsymbol{e}^\perp$ in the scenario, such that the two satisfy the operational identity $\boldsymbol{e}+\boldsymbol{e}^\perp = \boldsymbol{u}$.
More broadly, every measurement $M$ constitutes a set of effects that sums to the unit effect, giving the operational identity, 
\begin{equation}
    \sum_{\boldsymbol{e} \in M} \boldsymbol{e} = \boldsymbol{u};
\end{equation}
by linearity and the representation of the unit effect as the all-ones vector, then, it follows that 
\begin{equation} \label{mmtsOI}
    \sum_{\boldsymbol{e}\in M}\xi_{\boldsymbol{e}}(\lambda)=1
\end{equation}
for all $\lambda$ and  for every measurement $M$.

For transformations, one has linear equalities of the form
\begin{equation}
\left\{\sum_{\bf T \in \mathcal{T}} \alpha_{\bf T}^{(c)} { \bf T} = 0\right\}_c,
\label{eq:linearOpIdentityGPT}
\end{equation}
and the constraints on the ontological model from linearity are
\begin{equation} 
\forall \lambda,\lambda', c: \quad\sum_{\bf T \in \mathcal{T}} \alpha_{\bf T}^{(c)} \Gamma_{\bf T}(\lambda'|\lambda) = 0.
\label{eq:linearOpIdentityOM}
\end{equation}

Just as with states, one can efficiently find and enumerate all of these identities for effects and for transformations, as we show in Appendix~\ref{appendix:AllOpIdentitiesTransf}.

However, consideration of compositionality leads to new kinds of operational identities.
This is most obvious for transformations, where operational identities of the form 
\beq \label{seqconstr}
{\bf T} \circ {\bf T}' = { \bf T}''
\eeq
arise.
Even for states and effects, however, consideration of subsystem structure or composition with other processes leads to novel sorts of operational identities that are not of the linear forms above. These may involve sequential composition, parallel composition, partial traces of subsystems, combinations of several of these, and more. 
As one example, the fact that a state $\bomega_{S_{1}S_{2}}$ factorizes as $\bomega_{S_{1}S_{2}}:=\bomega_{S_1} \otimes \bomega_{S_2}$ can be written as a nonlinear operational identity
\begin{equation}
    \bomega_{S_1S_2} = \left[(\mathbb{1}_{S_1} \otimes {\bf u}_{S_2})\circ\bomega_{S_{12}}\right] \otimes \left[( {\bf u}_{S_1} \otimes\mathbb{1}_{S_1})\circ\bomega_{S_{12}}\right].
\end{equation}
Another example is given in Eq.~12 of Ref.~\cite{schmid2024addressing}, and a third example will be important later on in the manuscript (in Eq.~\eqref{GPTsource}). Even writing down a general form for an operational identity on a given type of process---much less giving an algorithm for finding a characterization of (a generating set of) all the operational identities for a set of such processes---is a difficult question that we leave open. 

In this work, we will focus on those operational identities of the linear forms above. 
The primary reason for this is because (as we will see) the constraints implied by such expressions for an ontological model are necessarily linear in the unknown quantifiers, and this leads to a tractable problem (specifically, to a linear program). In contrast, more general operational identities imply constraints on products of stochastic maps in the ontological representation, and consequently will involve nonlinearities on the unknowns---nonlinearities that we do not yet have techniques for addressing. 
A secondary reason is that it is only for operational identities of this form that we already have a general technique for characterizing all of the operational identities (the technique we introduce in Appendix~\ref{appendix:AllOpIdentitiesTransf}.)

Because we restrict attention to these `linear operational identities', we will be deriving necessary but not sufficient conditions for the existence of an ontological model. Equivalently, we find witnesses that are sufficient to certify nonclassicality, but not necessary. However, the inequalities from our program are sufficient as well as necessary for the existence of an ontological model that is linear on processes in the $\mathcal{PTM}$ scenario---it is just that the model need not satisfy all of the constraints from compositionality.\footnote{Indeed, some constraints from compositionality can even arise in a $\mathcal{PTM}$ scenario: one might have that ${\bf T}(\bomega) = \bomega'$ for some ${\bf T},\bomega,\bomega'$ in the scenario. Our program does not generically require that these operational identities be respected in the ontological model. (One can easily prove that they will be respected in any model of minimal ontic state space cardinality, but our program also does not require that the models be minimal dimensional.) }

\section{A naive extension of prior techniques}

In Ref.~\cite{Schmid2018}, a general algorithm was presented for deriving all of the noncontextuality inequalities for arbitrary prepare-measure experiments with respect to any fixed sets of operational equivalences among the preparations and among the measurements. We now revisit the central idea of that paper, which allows one to find these inequalities by solving the computational tasks of vertex enumeration and linear quantifier elimination.
Our presentation will moreover recast the results of Ref.~\cite{Schmid2018} into the current language of ontological models for GPTs, as opposed to noncontextual ontological models for (unquotiented) operational theories, in accordance with the discussion in Section~\ref{NCandGPTs}. 

We imagine a scenario consisting of a finite set of $N$ GPT states and a finite set of K GPT effects, although everything we prove would also apply to arbitrary sets of states and effects that had finitely many extremal vectors.

Consider a generic ontological model for the scenario with ontic state space $\Lambda$ (of potentially infinite cardinality), with epistemic states $\{\mu_{\bomega}\}_{\bomega}$, and with response functions $\{\xi_{{\bf e}}\}_{{\bf e}}$, reproducing the operational data in the scenario via
\begin{align}
{\bf e}\circ \bomega &= \int_{\Lambda} \xi_{{\bf e}}(\lambda) \mu_{\bomega}(\lambda) d\lambda.
\end{align}

Each ontic state $\lambda$ in any given ontological model assigns a probability to each of the $n$ effects, and so is associated with a vector $\Phi_{\lambda} \in \mathds{R}^\mathcal{E}$ defined component-wise by taking the $\mathbf{e}-$th component to be $ [\Phi_\lambda]_{\mathbf{e}} := \xi_\mathbf{e}(\lambda)$. If the effects in one's scenario are denoted $\mathbf{e}_1,...,\mathbf{e}_{K}$, then this vector is 
\begin{equation}
    \Phi_{\lambda} := \left(\xi_{{\bf e}_1}(\lambda),\xi_{\mathbf{e}_2}(\lambda),...,\xi_{\mathbf{e}_{K}}(\lambda)\right).
\end{equation}
We term such vectors {\em measurement assignments}.  
For a vector $\Phi_{\lambda}$ to be a possible measurement assignment, it must satisfy normalisation, positivity  and linearity, namely
\begin{align} 
&\quad [\Phi_\lambda]_{\mathbf{u}} = 1\label{mmtassign1}\\
\forall \mathbf{e}:  &\quad [\Phi_\lambda]_{\mathbf{e}} \ge 0 \label{mmtassign2} \\
\forall b: &\quad \sum_{\mathbf{e}\in\mathcal{E}} \alpha_\mathbf{e}^{(b)} [\Phi_\lambda]_{\mathbf{e}} 
= 0, \label{mmtassign3}
\end{align}
where the $\alpha_\mathbf{e}^{(b)}$ are defined based on the linear constraints holding among effects in the scenario (and  can be computed explicitly using the technique in Appendix~\ref{appendix:AllOpIdentitiesTransf}).

The set of logically possible measurement assignments are the vectors in $\mathds{R}^\mathcal{E}$ which satisfy these constraints. This forms a polytope within $\mathds{R}^\mathcal{E}$.  
To see this, first note that for any set of effects that sums to the unit effect, linearity implies that the representations of those effects satisfy $\sum_{{\bf e}\in M}\xi_{\bf e}(\lambda)=1$ for every $\lambda$. But every effect appears in at least one resolution of the unit effect (namely $\boldsymbol{e}+\boldsymbol{e}^\perp = \boldsymbol{u}$), so the fact that $\xi_{\bf e}(\lambda)\ge0$ holds for all $\lambda$ implies that  $\xi_\mathbf{e}(\lambda) \leq 1$ holds  for all $\lambda$.
Thus each measurement assignment lies inside the unit hypercube. Moreover, any operational identity among the effects implies a linear constraint on the components of $\Phi_{\lambda}$, and the vectors satisfying the linear constraint lie in a hyperplane. There are a finite number of these hyperplanes, as shown in Appendix~\ref{appendix:AllOpIdentitiesTransf}. The intersection of these finitely many hyperplanes with the unit cube gives a polytope.

This polytope captures the space of possible measurement assignments that could be made in any ontological model of the effects. One can find the vertices of this polytope---the convexly-extremal measurement assignments---by solving the computational task of vertex enumeration, as detailed in Ref.~\cite{Schmid2018}. We denote the vertices of the polytope by $\kappa' = 1,2,3,...$ and denote the measurement assignment made by vertex $\kappa'$ by $\widetilde{\Phi}_{\kappa'}$.  We have added the tilde here simply to denote that, once these quantities are found by vertex enumeration, they are numerically known quantities (while the prime is added for convenience of notation in later sections).  Therefore, the measurement assignment made by any given $\lambda$ can always be encoded in some vector $\Phi_\lambda$ lying inside this {\em measurement assignment polytope}, and this vector can always be written as a convex mixture 
\begin{equation} \label{eq:MAdecomp}
\Phi_\lambda = \sum_{\kappa'} w_\lambda(\kappa')\widetilde{\Phi}_{\kappa'} \end{equation} 
of the extremal measurement assignments, for some convex weights $\{w_\lambda(\kappa')\}_{\kappa'}$ satisfying $\sum_{\kappa'} w_\lambda(\kappa')=1$.

Writing just the $\mathbf{e}-$th component of Eq.~\eqref{eq:MAdecomp}, we obtain
\beq\label{decomp}
\forall \mathbf{e}, \lambda: \quad [\Phi_\lambda]_{\mathbf{e}} = \xi_\mathbf{e}(\lambda) = \sum_{\kappa'} w_\lambda(\kappa') [\widetilde{\Phi}_{\kappa'}]_{\mathbf{e}}.
\eeq
Consequently, we can write the empirical probabilities in the given ontological model as a finite sum, via
\begin{align}
{\bf e}\circ \bomega &= \int_{\Lambda} \xi_{{\bf e}}(\lambda) \mu_{\bomega}(\lambda) d\lambda \\
&= \int_{\Lambda} \Big( \sum_{\kappa'} [\widetilde{\Phi}_{\kappa'}]_{\mathbf{e}}w_\lambda(\kappa') \Big) \mu_{\bomega}(\lambda) d\lambda \\
&=  \sum_{\kappa'} [\widetilde{\Phi}_{\kappa'}]_{\mathbf{e}} \Big( \int_{\Lambda}w_\lambda(\kappa') \mu_{\bomega}(\lambda) d\lambda \Big) \\
&= \sum_{\kappa'} [\widetilde{\Phi}_{\kappa'}]_{\mathbf{e}} \nu_{\bomega}(\kappa'),\label{newprobs}
\end{align}
where we have defined $\nu_{\bomega}(\kappa') := \int_{\Lambda}  w_\lambda(\kappa') \mu_{\bomega}(\lambda) d\lambda$.

Since the numerical values for the components of the vectors $\{\widetilde{\Phi}_{\kappa'}\}_{\kappa'}$ can be computed explicitly, the operational probabilities in Eq.~\eqref{newprobs} are linear in the unknown quantities, namely $\{\nu_{\bomega}(\kappa')\}_{\kappa'}$. The problem of a GPT admitting an ontological model is then recast as a system of linear equations: those that arise by demanding the ontological model to reproduce the operational probabilities and those that arise by linear operational identities on states, together with normalization and positivity. By eliminating the unobserved quantities $\{\nu_{\bomega}(\kappa')\}_{\kappa'}$ from the system of equations using quantifier elimination, one can obtain a system of linear inequalities over the observed data alone, which form necessary and sufficient operational conditions for admission of an ontological model.

Let us now attempt to directly apply these methods to a $\mathcal{PTM}$ scenario.
Once again, one can explicitly compute the extremal measurement assignments, and then can
decompose generic measurement assignments into extremal ones, just as in Eqs.~\eqref{decomp}. Substituting this into the expression for the observable probabilities in the present scenario, one obtains
\begin{align}
&{\bf e}\circ  {\bf T} \circ \bomega = \int_{\Lambda,\Lambda'} \xi_{{\bf e}}(\lambda') \Gamma_{{\bf T}}(\lambda'|\lambda) \mu_{\bomega}(\lambda) d\lambda d\lambda' \nonumber \\
&= \int_{\Lambda,\Lambda'} \Big( \sum_{\kappa'} [\widetilde{\Phi}_{\kappa'}]_{\mathbf{e}}w_{\lambda'}(\kappa') \Big) \Gamma_{{\bf T}}(\lambda'|\lambda) \mu_{\bomega}(\lambda) d\lambda d\lambda' \nonumber \\
&=  \sum_{\kappa'} [\widetilde{\Phi}_{\kappa'}]_{\mathbf{e}}\int_{\Lambda} \Big( \int_{\Lambda'}w_{\lambda'}(\kappa') \Gamma_{{\bf T}}(\lambda'|\lambda) d\lambda' \Big)  \mu_{\bomega}(\lambda) d\lambda\nonumber \\
&=  \sum_{\kappa'} [\widetilde{\Phi}_{\kappa'}]_{\mathbf{e}}\int_{\Lambda} \tau_{{\bf T}}(\kappa'|\lambda) \mu_{\bomega}(\lambda) d\lambda, 
\end{align}
where $\tau_{{\bf T}}(\kappa'|\lambda) :=  \int_{\Lambda'}w_{\lambda'}(\kappa') \Gamma_{{\bf T}}(\lambda'|\lambda) d\lambda' $. 
Unfortunately, this expression still contains two sets of unknowns $\{\tau_{\bf T}(\kappa'|\lambda) \}_{
\bf T}$ and $\{\mu_{\bomega}(\lambda)\}_{\bomega}$, both of which are functions of a potentially continuous parameter $\lambda$; moreover, the expression is nonlinear in these unknowns. 

A natural resolution to these issues would present itself if we could decompose the epistemic states into convexly-extremal assignments in a manner analogous to what was done above for response functions. 

In direct analogy with the above, one might define a vector $\left(\mu_{\bomega_1}(\lambda),\mu_{\bomega_2}(\lambda),...,\mu_{\bomega_N}(\lambda)\right)$ of the weights assigned to a given $\lambda$ by each GPT state and then seek to characterize the set of possible such vectors. However, it is not clear how to make this most direct approach work, because it is not clear how to explicitly compute the extremal vectors of this sort, nor even that there are finitely many extremal vectors. 

Similar to what we had for measurement assignments, all such vectors live in the unit hypercube in  $\mathds{R}^\Omega$  and are subject to linear constraints due to the operational identities among states, which constitute hyperplanes slicing through the hypercube; valid vectors must live in the intersection of all of these. But while the set of measurement assignments was exactly the resulting polytope formed by the intersection of these equations, the epistemic states must satisfy additional constraints from normalization of epistemic states, and it is not clear how to incorporate these. In particular,  the constraint that $\sum_\lambda \mu_\omega(\lambda)=1$ for every $\omega$ is {\em not} a constraint on the components of such a vector, but rather is a constraint {\em among many such vectors}. Consequently, it does not simply define a constraint on the set of such vectors, and so it is not clear how (or if) it is possible to proceed in this direction.

Consequently, we do not see how to directly generalize the approaches from earlier work.
Rather, we will make progress by additionally drawing on the notion of flag-convexification~\cite{selby2023accessible,selby2023incompatibility} and by making an appropriate Bayesian inversion. 

Our approach initially followed a source\footnote{The term {\em source} refers to a preparation device that has a classical outcome as well as a GPT output.}-based approach like that in Ref.~\cite{Krishna_2017}, but ultimately we take a flag-convexification approach that is more general and more elegant. Indeed, when applied to the special case of prepare-measure scenarios, our approach is essentially the natural extension of the arguments in Ref.~\cite{Krishna_2017} to completely general scenarios (rather than to the specific sorts of sources studied therein).

\section{A linear program for computing $\mathcal{PTM}$ inequalities}

\begin{figure}[htb!]
\includegraphics[width=0.40\textwidth]{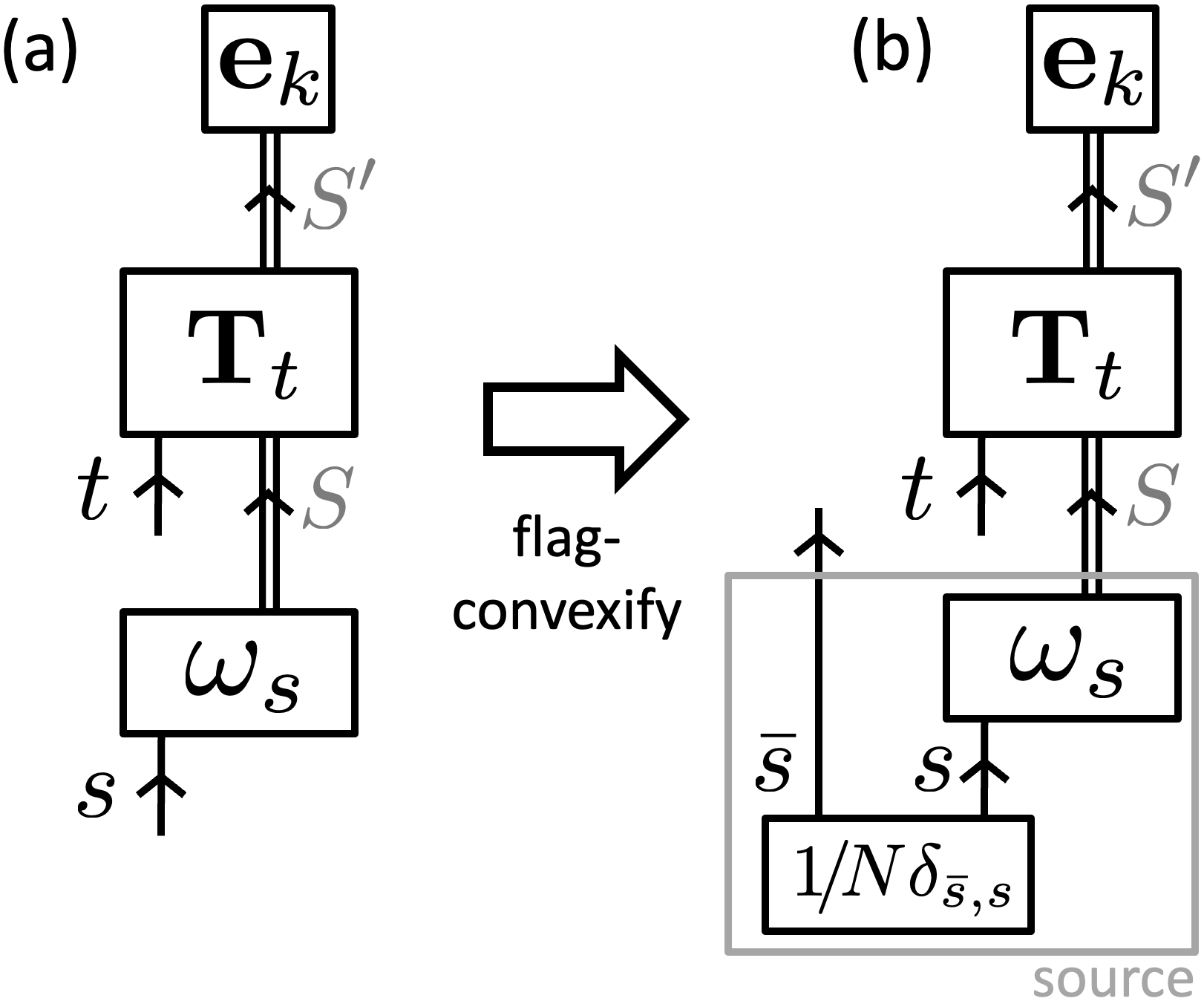}
\caption{(a) The scenario. (b) The flag-convexification of the scenario in (a). }\label{fcscenario}
\end{figure}

Say we have a set of states $\{\bomega_s\}_s$, a set of transformations $\{ {\bf T}_t\}_t$, and a set of effects $\{{\bf e}_k\}_k$, and we consider all possible combinations of a single state with a single transformation and a single effect, as in Figure~\ref{fcscenario}(a). 
Consider also the closely related {\em flag-convexified}~\cite{selby2023accessible} scenario shown in Figure~\ref{fcscenario}(b). In a flag-convexified scenario, one does not directly choose the setting variable $s$, but rather allows the setting's value to be selected probabilistically according to some fixed, full-support distribution over its possible values; one then copies the particular value obtained and treats it as an {\em outcome} variable (called a `flag') rather than an input setting. The specific flag-convexified scenario of interest to us takes the fixed distribution to be uniform over the $N$ possible setting values, so $p(s)=\frac1N$ and $p(s\bar{s})=\frac1N \delta_{s,\bar{s}}$. We denote the flag variable by $\bar{s}$. In the flag-convexified scenario, one has a single source (the object in the grey box in Figure~\ref{fcscenario}(b)) instead of a set of preparations.

In fact, one need not consider the flag-convexification procedure to be {\em physical}---i.e., something that one actually realizes by some probabilistic physical mechanism; rather, one can view this classical processing of the scenario as an inferential one, done in the mind of an agent. As the two scenarios differ only by an inferential processing of a classical variable, namely by Bayesian inversion, the empirical correlations, operational identites, noncontextuality inequalities, and ontological representation of processes in the two scenarios are related very simply.
For example, the empirical correlations $p(k|st)$ in the original scenario and the correlations $p(k\bar{s}|t)$ in the flag-convexified scenario are related by
\begin{align} \label{emprelation}
p(k\bar{s}|t) &= \sum_s p(k|st)p(s\bar{s})  \nonumber\\
&=\sum_s p(k|st)\frac1N\delta_{s\bar{s}}\nonumber \\
& =\frac1Np(k|s=\bar{s},t)
\end{align}

Clearly, then, 
\begin{equation}
\sum_{k,t,\bar{s}} \gamma_{k,\bar{s},t} p(k\bar{s}|t) + {\gamma_0}  \ge 0
\end{equation}
is a noncontextuality inequality for the flag-convexified scenario if and only if 
\begin{equation}
\sum_{k,t,s} \gamma_{k,s,t} p(k|st) + N\gamma_0 \ge 0
\end{equation}
is a noncontextuality inequality for the original scenario. 

Finally, if the representation of state $\bomega_s$ in the model is $\mu_s(\lambda)$, then the representation of the source is
\begin{equation}
p(\bar{s},\lambda)=\sum_s \frac1N\delta_{s\bar{s}}\mu_s(\lambda) = \frac1N\mu_{s=\bar{s}}(\lambda).
\label{RepresentationSources}
\end{equation}
(Here and henceforth, we will often simply use the subscript $s$ instead of the more explicit subscript $\bomega_s$, and similarly will use $k$ instead of ${\bf{e}}_k$, and $t$ instead of ${\bf{T}}_t$.) Equivalently, one has $p(\bar{s}=s,\lambda)=\frac1N\mu_s(\lambda)$, and so it follows that 
\begin{equation}\label{sourceopid}
\forall \lambda,a: \quad \sum_{s}\alpha^{(a)}_s p(\bar{s}=s,\lambda)=\frac1N \sum_{s}\alpha^{(a)}_s \mu_s(\lambda)=0,
\end{equation}
where we have used the constraints following from noncontextuality applied to operational identities among the states, namely (recalling Eq.~\eqref{OMIdentitiesStates})
\begin{equation}
\sum_{s} \alpha_{s}^{(a)} \mu_{s}(\lambda)=0.\label{OMidentitiesagain}
\end{equation}

In other words, mapping the set of states to a source as we have done (using the uniform distribution) amounts to just a constant rescaling of all of the states, which does not affect the linear relationships among them---i.e., the operational identities among them (both at the GPT level, and in the ontological model).

While all this may look trivial---we have merely been multiplying and dividing various objects by a constant factor, after all---we will see that deriving inequalities for the flag-convexified scenario is in fact easier than for the original scenario.

The above equations are all that we will need in order to derive inequalities for the flag-convexified scenario, and then to translate these back into inequalities for our original scenario. 
However, it is (tangentially) interesting to consider the source as a GPT object in its own right, and to think about what operational identities look like for this source.\footnote{ Elsewhere in this paper, we treat the classical systems as abstract objects about which one makes inferences, rather than as physical systems in their own right. In these paragraphs, we are essentially showing how this inferential view is consistent with viewing them as physical GPT systems.}
Naively, one might have expected that there are no operational identities that hold for the single source in our flag-convexified scenario, as there are no other sources with which to consider equivalences. But in fact, it is perfectly possible to have nontrivial operational identities relating to a single process, provided that that process has some substructure. This point has been made in Ref.~\cite{selby2023incompatibility} and in Section~III of Ref.~\cite{schmid2024addressing}; here, we see another example.

Since the GPT representation of state $s$ in the original scenario is $\bomega_s$, the GPT representation of the source in the flag-convexified scenario is
\begin{equation}
\frac1N \sum_s  |s)_{\bar{s}} \otimes \bomega_s,
\end{equation}
where $|s)_{\bar{s}}$ denotes the state of the classical GPT system $\bar{s}$ (here being viewed as an extremal state of a simplicial GPT) taking value $s$.
(We abuse notation slightly here by using $\bar{s}$ to denote a system, while at other times using it to denote a particular value of this classical system.) The source satisfies the operational identities 
\begin{equation}\label{GPTsource}
\sum_{s'} \alpha_{s'}^{(a)} \Big((s'| _{\bar{s}}\otimes \mathbb{1}_{S}\Big) \circ \left(\frac1N \sum_s  |s)_{\bar{s}} \otimes \bomega_s\right)=0,
 \end{equation}
one for each $a$ as in Eq.~\eqref{OpIdentitiesStates}.
Here, $ (s'|_{\bar{s}}$ is the effect on the simplicial system $\bar{s}$ that satisfies $ (s'|_{\bar{s}}\circ|s)_{\bar{s}}=\delta_{s,s'}$.  
Within any ontological model, the representation of Eq.~\eqref{GPTsource} is
\begin{align} 
0&=\sum_{s'} \alpha_{s'}^{(a)} \left(\frac1N \sum_s  \Big((s'|_{\Lambda_{\bar{s}}}\circ |s)_{\Lambda_{\bar{s}}}\Big) \otimes \mu_{s}(\lambda)\right)\\
&=\sum_s\alpha_s^{(a)} \mu_{s}(\lambda).
\end{align} 
Here, we have introduced the ontic state space $\Lambda_{\bar{s}}$ which has one ontic state for each of the possible GPT states $|s)_{\bar{s}}$, and we have introduced $(s'|_{\Lambda_{\bar{s}}}$, the response function representing effect $( s' |_{\bar{s}}$---namely, the function on $\Lambda_{\bar{s}}$ which is zero on all ontic states except the one associated with $|s')_{\Lambda_{\bar{s}}}$. 
This reproduces Eq.~\eqref{OMidentitiesagain}. 

We now proceed to derive the inequalities for the flag-convexified scenario.

Within an ontological model, the operational probabilities in the flag-convexified scenario must be equal to the composition of the stochastic maps representing the source, transformation, and effect, as
\begin{equation} \label{opfc2}
p(k\bar{s}|t) =  \int d\lambda d\lambda' \xi_{k}(\lambda') \Gamma_t(\lambda'|\lambda)\frac1N\mu_{\bar{s}}(\lambda),
\end{equation} 
where we have defined $\mu_{\bar{s}}(\lambda):=\mu_{s=\bar{s}}(\lambda)$. (Eq.~\eqref{opfc2} is consistent with Eqs.~\eqref{emprelation} and \eqref{RepresentationSources}, naturally.)
Noting that the marginal distribution over $\lambda$ induced by the source is $p(\lambda)=\sum_{\bar{s}}p(\bar{s},\lambda)$ and that $\mu_{\bar{s}}(\lambda)$ is simply another notation for a conditional probability distribution of the form $p(\lambda|\bar{s})$, we can Bayesian invert the latter to define
\begin{equation}\label{sourcebayesianinversion}
\xi_{\bar{s}}(\lambda):=p(\bar{s}|\lambda)= \frac{\mu_{\bar{s}}(\lambda)p(\bar{s})}{p(\lambda)}=\frac{\mu_{\bar{s}}(\lambda)}{Np(\lambda)}.
\end{equation}
As $\xi_{\bar{s}}(\lambda)$ is a conditional probability distribution over $\bar{s}$ given $\lambda$, it satisfies
\begin{equation}
\forall \lambda: \quad \sum_{\bar{s}}\xi_{\bar{s}}(\lambda)=1.
\end{equation}
We will refer to $\xi_{\bar{s}}(\lambda)$ as the {\em source response function} for state $\bar{s}$. It describes a retrodictive inference: the probability one would assign to outcome $\bar{s}$ of the source having occurred, given  that one knew that the ontic state output from the source was $\lambda$.

Substituting this into Eq.~\eqref{opfc2}, we have
\begin{equation} \label{opfc3}
p(k\bar{s}|t) =  \int d\lambda d\lambda' \xi_{k}(\lambda') \Gamma_t(\lambda'|\lambda)\xi_{\bar{s}}(\lambda)p(\lambda).
\end{equation}

It is here that one can understand our reasons for introducing the flag-convexified scenario. If we had not done this, then the distribution over $p(\lambda)$ in this expression would not be defined, as the distribution over $
\lambda$ would depend on the preparation. Ref.~\cite{Krishna_2017} addresses this problem by restricting attention to particular scenarios with special collections of sources which all generate the same distribution over $\lambda$ when their outcomes are marginalized over; here, we need no such restriction, as the procedure of flag-convexification always leads to a single source, for which there is {\em trivially} a unique distribution $p(\lambda)$ when its outcome is marginalized over. (A secondary advantage of our approach is that it involves considerably less bookkeeping.)

We now follow Ref.~\cite{Krishna_2017} in convexly decomposing the source response function $\xi_{\bar{s}}(\lambda)$ in a manner similar to what was done for measurement response functions. 

Each ontic state $\lambda$ in the ontological model assigns a (retrodictive) probability to each of the $N$ possible outcomes of the source, and so is associated with a vector 
\begin{equation}
\Psi_\lambda := \left(\xi_{\bar{s}=1}(\lambda),\xi_{\bar{s}=2}(\lambda),...,\xi_{\bar{s}=N}(\lambda)\right)
\end{equation}
indexed by the states, which we term a {\em source assignment}. 
Since each element of this vector is defined as $\xi_{\bar{s}}(\lambda)=\frac{\mu_{\bar{s}}(\lambda)}{Np(\lambda)}$, Eq.~\eqref{sourceopid} places constraints on this vector, namely, for all $\lambda$:
\begin{equation} \label{sourceopid2}
\sum_{\bar{s}}\alpha^{(a)}_{\bar{s}} \xi_{\bar{s}}(\lambda)=\sum_{\bar{s}}\alpha^{(a)}_{\bar{s}} \frac{\mu_{\bar{s}}(\lambda)}{Np(\lambda)}=\frac{0}{Np(\lambda)}=0.
\end{equation}
(Note that $p(\lambda)\neq0$ for all $\lambda$, since the support of $p(\lambda)$ is the union of the supports of all the epistemic states, and any nontrivial $\lambda$ is in the support of at least one epistemic state.)  

Much in analogy to Eqs.~\eqref{mmtassign1}-\eqref{mmtassign3} for effects, a vector $\Psi_{\lambda}$ is a possible source assignment if and only if it satisfies 
\begin{align} 
&\quad \sum_{\bar{s}}[\Psi_\lambda]_{\bar{s}} = 
1;\label{sourceassign1}\\
\forall \bar{s}:  &\quad [\Psi_\lambda]_{\bar{s}} \ge 0 \label{sourceassign2}; \\
\forall a: &\quad \sum_{\bar{s}} \alpha_{\bar{s}}^{(a)} [\Psi_\lambda]_{\bar{s}} 
= 0, \label{sourceassign3}
\end{align}
where the $\alpha_{\bar{s}}^{(a)}$ are defined based on the linear constraints holding among effects in the scenario (and  can be computed explicitly using the technique in Appendix~\ref{appendix:AllOpIdentitiesTransf}).

The set of all such assignments forms a polytope, whose vertices we label by $\kappa$, with corresponding extremal source assignments $\widetilde{\Psi}_\kappa$. 
Consequently, we can decompose any given assignment into extremal source assignments as 
\begin{equation}
\Psi_\lambda = \sum_{\kappa} v_\lambda(\kappa)\widetilde{\Psi}_\kappa
\end{equation} 
for some convex weights $\{ v_\lambda(\kappa)\}_{\kappa}$. Component-wise, one has
\beq
\forall \bar{s},\lambda: \quad [\Psi_{\lambda}]_{\bar{s}} =  \xi_{\bar{s}}(\lambda) = \sum_{\kappa} v_\lambda(\kappa) [\widetilde{\Psi}_\kappa]_{\bar{s}},\label{Sdecomp}
\eeq

And as in Eq.~\eqref{decomp}, we also decompose the measurement response functions into extremal measurement assignments:
\beq \label{decomp2}
\forall k,\lambda: \quad [\Phi_\lambda]_{k} = \xi_k(\lambda) = \sum_{\kappa'} w_\lambda(\kappa')[\widetilde{\Phi}_{\kappa'}]_{k}.
\eeq

For both the source and for the effects, the extremal assignments can be found explicitly by vertex enumeration, after which they are numerically {\em known} quantities (a fact which we emphasize throughout by denoting them with a tilde).

Substituting Eq.~\eqref{Sdecomp} and Eq.~\eqref{decomp2} into Eq.~\eqref{opfc3}, we obtain
\begin{align} \label{opfc4}
&p(k\bar{s}|t) \\ \nonumber
&=  \int d\lambda d\lambda' \sum_{\kappa',\kappa} w_{\lambda'}(\kappa')[\widetilde{\Phi}_{\kappa'}]_{k} \Gamma_t(\lambda'|\lambda) v_\lambda(\kappa) [\widetilde{\Psi}_\kappa]_{\bar{s}}p(\lambda).
\end{align}

Reordering the summations, we obtain
\begin{align} \label{reordering}
&p(k\bar{s}|t) \\
&=   \sum_{\kappa,\kappa'} [\widetilde{\Phi}_{\kappa'}]_{k} [\widetilde{\Psi}_{\kappa}]_{s}  \int d\lambda d\lambda' w_{\lambda'}(\kappa') \Gamma_t(\lambda'|\lambda) v_\lambda(\kappa) p(\lambda). \nonumber
\end{align}

Next, we define 
\begin{equation} \label{defnPT}
p(\kappa'\kappa|t):=\int d\lambda d\lambda' w_{\lambda'}(\kappa') \Gamma_t(\lambda'|\lambda)v_\lambda(\kappa) p(\lambda),
\end{equation}
a conditional probability distribution which (as one can check directly) satisfies 
\begin{equation}\label{causalindep}
p(\kappa|t) \equiv \sum_{\kappa'} p(\kappa' \kappa|t) = p(\kappa),
\end{equation} 
or equivalently,  
\begin{equation}
\forall t,t': \quad \sum_{\kappa'}\big( p(\kappa' \kappa|t) - p(\kappa' \kappa|t')\big) =0.
\end{equation}
In other words, learning $t$ does not provide information about $\kappa$. (Intuitively, this is a consequence of the fact that $\kappa$ is prepared by the source temporally prior to the transformation ${ \bf T}_t$ in the experiment.)
 
Definition~\ref{defnPT} allows us to write Eq.~\eqref{reordering} as
\begin{align} \label{empdatab}
p(k\bar{s}|t) =   \sum_{\kappa,\kappa'} [\widetilde{\Phi}_{\kappa'}]_{k} [\widetilde{\Psi}_\kappa]_{\bar{s}}p(\kappa' \kappa|t).
\end{align}

The linear operational identities among transformations (Eq.~\eqref{eq:linearOpIdentityGPT}) imply that in the ontological model, one has
\beq \label{tnc1}
\forall c: \quad \sum_t \alpha_t^{(c)} \Gamma_t(\lambda'|\lambda) = 0.
\eeq
By Eq.~\eqref{defnPT}, $p(\kappa' \kappa|t)$ similarly satisfies
\begin{equation} \label{TNCnewmodelBI}
\forall c: \quad \sum_t \alpha_t^{(c)} p(\kappa' \kappa|t) = 0,
\end{equation}
which is proven by noting that
\begin{align} 
&\sum_t \alpha^{(c)}_t 
p(\kappa' \kappa|t) \\ \nonumber
&=\sum_t \alpha^{(c)}_t\int d\lambda d\lambda' w_{\lambda'}(\kappa') \Gamma(\lambda'|\lambda,t)v_\lambda(\kappa) p(\lambda) \\ \nonumber
&=\int d\lambda d\lambda' w_{\lambda'}(\kappa') \Big(\sum_t \alpha^{(c)}_t\Gamma(\lambda'|\lambda,t)\Big)v_\lambda(\kappa) p(\lambda) \\ \nonumber
 &=0.
\end{align}

At this point, we have reduced the problem to one with a finite set of unknowns, $\{ p(\kappa' \kappa|t)\}_{\kappa',\kappa,t}$, which appear linearly in the expression (Eq.~\eqref{empdatab}) for the empirical operational data for the flag-convexified scenario.\footnote{Although we consider it extremely likely, we have not actually {\em proven} that more general operational identities involving compositions of transformations (like those in Eq.~\eqref{seqconstr}, as just one example) lead to nonlinearities in the unknown quantities once transformations are represented by probability distributions $p(\kappa' \kappa|t)$ rather than conditionals like $\Gamma(\lambda'|\lambda)$. Answering this question would require finding an appropriate notion of composition for two probability distributions $p(\kappa' \kappa|t)$ and $p(\kappa'' \kappa'|t')$, which is an open question. 
} 

In summary, we have shown that
a necessary condition for the flag-convexified $\mathcal{PTM}$ scenario to admit of an ontological model is that
\vspace{1.5pt}\begin{align*}
\exists  \{p(\kappa' \kappa|t)\}_{\kappa',\kappa,t} \text{  such that } \nonumber
\end{align*}\vspace{-2pt}\begin{alignat}{2}
\forall \kappa',\kappa,t: \quad&& \quad  p(\kappa' \kappa|t) &\ge 0,  \label{mc0fc}\\[4pt] 
\forall t: \quad&& \quad \sum_{\kappa',\kappa} p(\kappa' \kappa|t) &= 1\,, \label{mc1fc} \\ 
\forall \kappa,t,t': \quad&& \quad \sum_{\kappa'}\big( p(\kappa' \kappa|t) - p(\kappa' \kappa|t')\big) &=0, \label{causalindep2fc} \\ 
\forall \kappa',\kappa, c:\quad&& \quad \sum_t \alpha_t^{(c)} p(\kappa' \kappa|t) &= 0, \label{mc2fc} 
\end{alignat}\begin{flalign}
\forall k,\bar{s},t: \ \   \sum_{\kappa,\kappa'} [\widetilde{\Phi}_{\kappa'}]_{k} [\widetilde{\Psi}_\kappa]_{\bar{s}}p(\kappa'\kappa|t) &= p(k\bar{s}|t), \label{mc3fc}
\end{flalign}\normalsize
\noindent 
where $\{\widetilde{\Phi}_{\kappa'}\}_{\kappa'}$ is the set of vertices of the measurement-assignment polytope (defined by Eqs.~\eqref{mmtassign1}-\eqref{mmtassign3}),  $\{\widetilde{\Psi}_\kappa\}_{\kappa}$ is the set of vertices of the source-assignment polytope (defined by Eqs.~\eqref{sourceassign1}-\eqref{sourceassign3}),  
and where $t,t'$ range over the finite set of transformations in the scenario, $\bar{s}$ the finite set of source outcomes, and $k$ the finite set of effects in the scenario.

To obtain constraints that refer only to operational probabilities, one must eliminate the unobserved $\{p(\kappa' \kappa | t)\}_{\kappa',\kappa,t}$ from the system of equations~\eqref{mc0fc}-\eqref{mc3fc}, obtaining a system of linear inequalities over the $\{p(k\bar{s}|t)\}_{k,\bar{s},t}$ alone. 

The standard method for solving this problem of \textit{linear quantifier elimination} is the Chernikov-refined Fourier-Motzkin algorithm~\cite{DantzigEaves,BalasProjectionCone,jones2004equality,Shapot2012,Bastrakov2015}, which is implemented in a variety of software packages (some of which can be found listed in Ref~\cite{Schmid2018}).

By performing quantifier elimination, one obtains a list of linear inequalities over the observed data, which we denote by $\mathcal{H}:=\{h^1, h^2,...\}$;  we denote the coefficients of a specific facet inequality $h$ by $\gamma^{(h)}_{k,\bar{s},t}$, and denote the constant term in that inequality by $\gamma^{(h)}_0$. In other words, a necessary condition for the flag-convexified $\mathcal{PTM}$ scenario to admit of an ontological model is that
\begin{align} \label{c3}
\forall h\in\mathcal{H}\,:\quad\sum_{k,\bar{s},t} \gamma^{(h)}_{k,\bar{s},t} \ p(k\bar{s}|t) + \gamma^{(h)}_0 \ge 0\,,
\end{align}
where $\mathcal{H}$ is the set of $n$ inequalities. 

\sloppy Recalling the relationships between the two scenarios (for example, Equation~\eqref{emprelation}, which tells us that ${p(k|s=\bar{s},t)=Np(k\bar{s}|t)}$), we can now link these inequalities for the flag-convexified scenario back to the inequalities for the original scenario.

In particular, it follows that a necessary condition for a $\mathcal{PTM}$ scenario to admit of an ontological model is that
\begin{align} \label{c4}
\forall h\in\mathcal{H}\,:\quad \sum_{k,s, t} \gamma^{(h)}_{k,s,t} \ p(k|st) + N\gamma^{(h)}_0 \ge 0\,,
\end{align} 
where $N$ 
is the number of GPT states in the set $\mathcal{P}$, and where $\mathcal{H}$ is the set of inequalities resulting from eliminating all free parameters $\{p(\kappa' \kappa |t)\}_{\kappa',\kappa,t}$ in the system of equations in Eqs.~\eqref{mc0fc}-\eqref{mc3fc}

Finally, we write down a linear program that directly solves for the inequalities of interest, namely those in Eq.~\eqref{c4}, by making a small modification to the linear program above---namely by changing a factor of $N$ in Eq.~\eqref{mc3fc}. The proof that this is the correct modification is given in Appendix~\ref{appendix:mapLPs}.

\setlength{\belowdisplayskip}{0pt plus 0pt}
\setlength{\abovedisplayskip}{0pt plus 0pt}
\setlength\abovedisplayshortskip{0pt}
\setlength\belowdisplayshortskip{0pt}
\begin{samepage}
\begin{formulation}  \label{c2} 
A necessary condition for the $\mathcal{PTM}$ scenario defined by sets of GPT processes $\mathcal{P},\mathcal{T},$ and $\mathcal{M}$ to admit of an ontological model is that
\vspace{1.5pt}\begin{align*}
\exists  \{p(\kappa' \kappa|t)\}_{\kappa',\kappa,t} \text{  such that } \nonumber
\end{align*}\vspace{-2pt}\begin{alignat}{2}
\forall \kappa',\kappa,t: \quad&& \quad  p(\kappa' \kappa|t) &\ge 0,  \label{mc0}\\[4pt] 
\forall t: \quad&& \quad \sum_{\kappa',\kappa} p(\kappa' \kappa|t) &= 1\,, \label{mc1} \\ 
\forall \kappa,t,t': \quad&& \quad \sum_{\kappa'}\big( p(\kappa' \kappa|t) - p(\kappa' \kappa|t')\big) &=0, \label{causalindep2} \\ 
\forall \kappa',\kappa, c:\quad&& \quad \sum_t \alpha_t^{(c)} p(\kappa' \kappa|t) &= 0,
 \label{mc2} 
\end{alignat}\begin{flalign}
\forall k,s,t: \ \   N\sum_{\kappa,\kappa'} [\widetilde{\Phi}_{\kappa'}]_{k} [\widetilde{\Psi}_\kappa]_{\bar{s}}p(\kappa'\kappa|t) &= p(k|s=\bar{s}t), \label{mc3}
\end{flalign}\normalsize 
where $\{\widetilde{\Phi}_{\kappa'}\}_{\kappa'}$ is the set of vertices of the measurement-assignment polytope (defined by Eqs.~\eqref{mmtassign1}-\eqref{mmtassign3}) and $\{\widetilde{\Psi}_\kappa\}_{\kappa}$ is the set of vertices of the source-assignment polytope (defined by Eqs.~\eqref{sourceassign1}-\eqref{sourceassign3}). 
\end{formulation}
\end{samepage}\normalsize

This is our first main result: a linear program that gives noncontextuality inequalities for a $\mathcal{PTM}$ scenario to admit of an ontological model.
These inequalities take into account the constraints from all linear operational identities. 

However, they do not take into account constraints arising from operational identities involving general compositionality (such as sequences of transformations or subsystem structure), nor the constraint that the identity GPT transformation should be represented as the identity on the ontic state space, nor the constraint (from diagram-preservation) that isomorphic GPT systems should be associated with isomorphic ontic state spaces. Thus, tighter inequalities will generally be possible.

Note that the linear program does not require as input any explicit specification of the GPT processes, but it does require one to specify the exact (linear) operational identities that these processes satisfy. Of course, determining these operational identities in the first place is most easily done if one begins with their explicit specification as GPT processes, in which case one can find them using the technique in Appendix~\ref{appendix:AllOpIdentitiesTransf}. And accurately determining their GPT representation requires assumptions of tomographic completeness, as expounded in Refs.~\cite{daley2021experimentally,grabowecky2021experimentally,mazurek2021experimentally}. However, our program itself does not require that the sets of states, effects, and transformations be tomographic relative to one another. 

Moreover, one can include an arbitrary subset of valid operational identities as input to the program, and for any subset one will still obtain valid noncontextuality inequalities; however, the more one includes, the tighter these inequalities will become.

One way to visualize the linear program \ref{c2} is to recast it in matrix form:
\begin{align}
\label{eq:programCheckExistence}
    \exists\, &\bs{x}\,\text{such that,}\nonumber\\
    \mathds{M}\cdot &\bs{x} = \bs{b},\\
    \text{and}\, &\bs{x}\geq 0\nonumber,
\end{align}
\sloppy where $\bs{x}$ contains the unknowns $\{p(\kappa',\kappa|t)\}_{\kappa',\kappa,t}$ and the rows of matrix $\mathds{M}$ carry the coefficients in the summations on the left-hand side of Eqs.~\eqref{mc1}-\eqref{mc3}; the entries of vector $\bs{b}$  contain the probabilities $\{p(k|s,t)\}_{k,s,t}$ as well as some zeros and ones---the zeros coming from Eqs.~\eqref{causalindep2} and~\eqref{mc2} and the ones coming from the normalization conditions in Eqs.~\eqref{mc1}. That is, the $\mathcal{PTM}$ scenario of interest fixes the matrix $\mathds{M}$ (through the extremal measurement assignments and extremal source assignments that one computes explicitly and through the operational identities on transformations); the noncontextuality inequalities are the necessary and sufficient conditions on the entries of the vector $\bs{b}$ for the system of equations to have a solution (i.e., to guarantee the existence of such a  vector $\bs{x}$).

Thus far, we have discussed applying linear quantifier elimination to the system of equations comprising this linear program in order to get a complete set of inequalities. In the next section, we discuss how if one has a specific data table $\{p^*(k|st)\}_{k,s,t}$, one can simply evaluate the linear program directly to test if that data admits of a noncontextual model.

\subsection{Efficiently solving the decision problem: `Does my data table admit of a noncontextual model?'}

For the purposes of witnessing genuine nonclassicality in a specific given experiment, one need not derive all the noncontextuality inequalities for a {\em generic} scenario with the relevant operational identities. Much as in Ref.~\cite{Schmid2018}, we now give a linear program to directly test whether a set of {\em numerical} data admits of a noncontextual explanation or not. If the data does not admit of a noncontextual model, then the linear program returns the inequality which is maximally violated by the data, which serves as a witness of the failure of noncontextuality. If it does admit of one, the program returns such a model. Furthermore, this test is more computationally efficient than deriving all the inequalities for a generic scenario and then testing if these are satisfied. 

Let us see explicitly how this can be derived as an instance of our linear program~\ref{c2}. Consider the matrix form of Eq.~\eqref{eq:programCheckExistence}, but now we treat $\bs{b}$ not as a variable, but as a known quantity (which we denote as $\bs{b^*}$):
\begin{align}
\label{eq:programCheckExistence2}
    \exists\, &\bs{x}\,\text{such that,}\nonumber\\
    \mathds{M}\cdot &\bs{x} = \bs{b^*},\\
    \text{and}\, &\bs{x}\geq 0\nonumber.
\end{align}
This program is feasible if and only if a noncontextual model for the numerical data table exists. One can thus check for existence of such an $\bs{x}$ by optimizing a constant objective function, so the program will return any vector $\bs{x}$ satisfying the constraints. If this is infeasible, then one gets a proof of contextuality. 

Furthermore, one can have a certificate of infeasibility through the program dual to that in Eq.~\eqref{eq:programCheckExistence2}, called the Farkas dual. This is a consequence of the fact that the optimisation problem
\begin{align}
\label{eq: DualCheckingExistence}
    \min_{y}\, &\bs{y}\cdot \bs{b^*},\,\text{such that}\nonumber\\
    \bs{1}\,\geq &\bs{y}\cdot \mathds{M}\geq 0,
\end{align}
has a solution $\bs{y}\cdot \bs{b^*}\geq 0$ if and only if the primal program in Eq.~\eqref{eq:programCheckExistence} is feasible. Therefore, if the program in Eq.~\eqref{eq: DualCheckingExistence} returns $\bs{y}\cdot\bs{b^*}<0$, one has a certificate that the primal program is infeasible and, therefore, that the data cannot be explained classically. 

In this way, the Farkas dual returns a certificate of nonclassicality in the form of the vector $\bs{y}$, which via the dot product with $\bs{b^*}$, corresponds to a noncontextuality inequality that is violated by the data table $\{p^*(k|s,t)\}_{k,s,t}$: the coefficients of the inequality being those elements of $\bs{y}$ that multiply the entries of $\bs{b^*}$ containing the empirical probabilities, and the noncontextual bound being the sum of all elements of $\bs{y}$ that multiply the normalization terms.
Therefore, our linear program in Eqs.~\eqref{mc0}-\eqref{mc3} gives us a direct witness of nonclassicality for one's empirical data. 

Moreover, the minimization in Eq.~\eqref{eq: DualCheckingExistence} guarantees that there are no other inequalities defined by a vector $\bs{y}$ satisfying $\bs{1}\geq y\cdot \mathds{M}\geq 0$ that lead to a higher violation by the data. In this sense, the program returns the inequality that is maximally violated by one's data.

\subsection{Constructing a candidate ontological model from the program's output}

Provided an ontological model exists for one's scenario, the linear program will output the quantities $\{p(\kappa' \kappa|t)\}_{\kappa,\kappa',t}$.  From vertex enumeration, one has the vectors $\{\widetilde{\Psi}_\kappa\}_\kappa$  and $\{\widetilde{\Phi}_{\kappa'}\}_{\kappa'}$. From these, one can construct a candidate explicit ontological model for the scenario, as follows.

First, one takes the ontic state space for system $S$ to be the set of all $\kappa$ for which $\sum_{\kappa'}p(\kappa,\kappa')>0$. (This condition picks out only those $\kappa$'s that are in the support of at least one epistemic state, as one can see from Eq.~\eqref{eq:epistemicStateOMfromLP}.)
We take the ontic state space of $S'$ to be the set of all $\kappa'$. The representation $\xi_k(\kappa')$ of effect ${\bf e}_k$ in the model is simply 
\begin{equation}
\xi_k(\kappa'):=[\widetilde{\Phi}_{\kappa'}]_{k}.
\end{equation}
 The representation of transformation ${\bf T}_t$ is taken to be
\begin{equation}
\Gamma_t(\kappa'|\kappa) := \frac{p(\kappa' \kappa|t)}{p(\kappa)},
\end{equation}
where
$p(\kappa) = \sum_{\kappa'} p(\kappa' \kappa|t)$ is independent of $t$, as in Eq.~\eqref{causalindep}.
Finally, the representation of state $\bomega_s$ is taken to be
\begin{equation}
\label{eq:epistemicStateOMfromLP}
\mu_{s}(\kappa):=[\tilde{\Psi}_\kappa]_{\bar{s}=s}\frac{p(\kappa)}{p(\bar{s}=s)}=N[\tilde{\Psi}_\kappa]_{\bar{s}=s}p(\kappa).
\end{equation} 

Intuitively, this model is simply what one gets when undoing the Bayesian inversion used from Eq.~\eqref{sourcebayesianinversion} onwards. One can also check explicitly that this is a valid ontological model. The representations have the appropriate form; e.g., $\mu_s(\kappa)$ is indeed a probability distribution, since it is positive and since 
\begin{equation}
\sum_\kappa \mu_s(\kappa) = \sum_\kappa N[\widetilde{\Psi}_{\kappa}]_{s}p(\kappa) =1.
\label{eq:NormalizedMu}
\end{equation}
The last equality follows from Eq.~\eqref{mc3} and Eq.~\eqref{mmtassign1}. In particular, by summing both sides of Eq.~\eqref{mc3} over all effects ${\bf e}_k$ in any single measurement $M$, we get 
\begin{align}
& \sum_{{\bf e}_k\in M}\left[N\sum_{\kappa',\kappa}[\tilde{\Phi}_{\kappa'}]_k[\tilde{\Psi}_\kappa]_s  p(\kappa,\kappa'|t)\right] \nonumber \\
&= \sum_{{\bf e}_k\in M} p(k|st)=1.
\end{align}
Recalling Eq.~\eqref{mmtsOI}, we then have
\begin{align}
    \hspace*{-1cm}&1= N\!\sum_{\kappa',\kappa}\!\!\left[\sum_{{\bf e}_k\in M}[\tilde{\Phi}_{\kappa'}]_k\!\right]\![\tilde{\Psi}_{\kappa}]_sp(\kappa,\kappa'|t) \nonumber \\
    &= N\!\sum_{\kappa',\kappa}[\tilde{\Psi}_{\kappa}]_sp(\kappa,\kappa'|t)\nonumber\\
    &=N\sum_\kappa [\tilde{\Psi}_{\kappa}]_s\sum_{\kappa'}p(\kappa',\kappa|t)\nonumber \\    
    &= N\sum_\kappa [\tilde{\Psi}_{\kappa}]_sp(\kappa) ,
\end{align}
which proves Eq.~\eqref{eq:NormalizedMu}.
The model also reproduces the empirical data, since
\begin{align} \label{empdatab2}
 &   \sum_{\kappa,\kappa'} \xi_k(\kappa') \Gamma(\kappa'|\kappa,t)\mu_s(\kappa)\\ 
&=   \sum_{\kappa,\kappa'} [\widetilde{\Phi}_{\kappa'}]_{k} \frac{p(\kappa' \kappa|t)}{p(\kappa)} N[\widetilde{\Psi}_{\kappa}]_{s}p(\kappa)\\
&= N\sum_{\kappa,\kappa'} [\widetilde{\Phi}_{\kappa'}]_{k} [\widetilde{\Psi}_{\kappa}]_{s}p(\kappa'\kappa|t) = p(k|st),
\end{align}
where the last equality is from Eq.~\eqref{mc3}.
By construction, the representation of effects in the model respects all the operational identities among effects. Those holding for states are also satisfied, since for every $a$,
\begin{align}
\sum_s \alpha_s^{(a)} \mu_s(\kappa) &= \sum_s \alpha_s^{(a)} N[\tilde{\Psi}_{\kappa}]_sp(\kappa) \nonumber \\ &= N p(\kappa)\sum_{s} \alpha_{s}^{(a)}[\tilde{\Psi}_{\kappa}]_s=0,
\end{align}
 where the last equality follows from Eq.~\eqref{sourceassign3}. 
The model also respects the linear operational identities for transformations, since for every $c$,
\begin{align}
\sum_t \alpha_t^{(c)} \Gamma_t(\kappa'|\kappa) &= \sum_t \alpha_t^{(c)} \frac{p_t(\kappa',\kappa)}{p(\kappa)} \\
&= \frac{1}{p(\kappa)}\sum_t \alpha_t^{(c)} p_t(\kappa',\kappa) = 0, 
\end{align}
where the last equality follows from Eq.~\eqref{TNCnewmodelBI}.

This construction does not necessarily satisfy all possible constraints to be a valid ontological model of the GPT in question, since our linear program does not enforce all possible constraints from diagram-preservation. For example, it may not represent the identity transformation by the identity. However, one can check explicitly if all relevant constraints are indeed satisfied, in which case one has an ontological model for the scenario. (It is usually not important to construct explicit ontological models; the main reason one might wish to do so is simply as an {\em explicit check} that one's scenario is in fact classically explainable.)

\section{Example: the single-qubit stabilizer}
\label{sec:stabqubit}

We now give an example of how our linear program can be used to derive noncontextuality inequalities for transformations, by applying it to a fragment of the stabilizer subtheory~\cite{gottesman1997stabilizer,gottesman1998heisenberg}.  The stabilizer subtheory has an important role in quantum computation, as it can be implemented in a fault-tolerant manner, (though it can be made universal only through the use of non-stabilizer `magic' states). This theory is known to be nonclassical in all even dimensions, and classical in all odd dimensions~\cite{Schmid2022Stabilizer}. The single-qubit stabilizer fragment in particular has the interesting feature that it admits of a noncontextual model for the prepare-measure scenario, but not for the prepare-transform-measure scenario~\cite{Lillystone2019}. 

Consider the fragment of the stabilizer subtheory containing all single-qubit stabilizer states and effects, together with the set of stabilizer transformations $\mathcal{T}=\{I,\mathcal{Z},S,S^{-1}\}$, where $I$ is the identity, $\mathcal{Z}$ is the $Z$-gate transformation, and $S$ is the phase gate.
Recall that the set of stabilizer states on a single qubit includes all and only eigenstates of the Pauli observables $X,Y,$ and $Z$, and the set of stabilizer effects includes all and only the projectors onto these states. 

The transformations in $\mathcal{T}$ satisfy the operational identity
\begin{align}
    \frac{1}{2}\left(I(\cdot)+\mathcal{Z}(\cdot)\right) = \frac{1}{2}\left(S(\cdot) + S^{-1}(\cdot)\right).
    \label{eq:OperIdTransfStabilizer}
\end{align} 
Using the technique of Appendix~\ref{appendix:AllOpIdentitiesTransf}, one can show that this is in fact the only operational identity for the scenario (or more precisely, the only generating operational identity).
The only operational identities among the effects are
\begin{align}
    \frac12\!\ketbra{1}{1}+\frac12\!\ketbra{0}{0} &=\frac12\!\ketbra{+}{+}+\frac12\!\ketbra{-}{-} \nonumber \\
    &= \frac12\!\ketbra{+i}{+i}+\frac12\!\ketbra{-i}{-i},
\end{align}
and the only operational identities among the states are of exactly the same form. That these are the unique (generating) identities can again be checked using the techniques of Appendix~\ref{appendix:AllOpIdentitiesTransf}.

Given these, one can characterize the source-assignment and measurement-assignment polytopes. For the measurement assignment polytope, vertex enumeration yields the eight deterministic assignments to the three binary outcomes. (This is consistent with the result of Ref.~\cite{Schmid2018} that whenever there are no operational identities other than normalization, the extremal assignments are all and only the deterministic ones.) For the source-assignment polytope, vertex enumeration yields the same eight deterministic assignments, but where each is multiplied by $\frac13$.
Because each of $\kappa$ and $\kappa'$ range over 8 possibilities and $t$ ranges over 4 possibilities, $p(\kappa,\kappa'|t)$  is a vector with $256$ entries. Furthermore, one can check that $\mathds{M}$ is a $260\times 256$ matrix capturing the restrictions\footnote{Note that the number of rows in $\mathbb{M}$ (in this case 260), is the sum of the number of restrictions implied by each of Eqs.~\eqref{mc1}-\eqref{mc3}. The number of restrictions implied by Eq.~\eqref{causalindep2} is $|T|(|T|-1)/2$ (where $|T|$ is the number of transformations in the scenario), since one counts only pairs of transformations $T$ and $T'$ that are distinct, and the order is irrelevant.  The number of restrictions implied by the other equations is the product of the cardinalities of the sets of variables involved.} in Eqs.~\eqref{mc1}-\eqref{mc3}.
Using the dual linear program in Eq.~\eqref{eq: DualCheckingExistence}, one can determine whether or not one can reproduce the quantum predictions $\bs{e}_k\circ\bs{T}_t\circ\bs{\omega}_s$ (which of course are given by the Hilbert-Schmidt inner product ${\rm Tr}[\bs{e}_{k}{\bf T}_t(\bs{\omega}_{s})]$) for this fragment of quantum theory in some noncontextual ontological model. As it turns out, this is not feasible, and a certificate of infeasibility can be obtained via the dual program. Defining the shorthand $p_{stk}:=\bs{e}_k\circ {\bf T}_t\circ\bomega_s$, the program gives a noncontextuality inequality,
\begin{align}
&    {\rm \beta} 
:= \\ 
 & 3\left[p_{{121}} \!-\! p_{{212}} \!+\! p_{{644}}  \!+\! p_{{234}} \!+\!
    p_{{335}} \!-\! p_{{445}} \!-\! p_{{146}} \!+\! p_{{326}} \!+\! p_{{416}}\right]\nonumber \\
    &+2\left[
    -p_{{311}}
    +p_{{322}}
    -p_{{333}}
    +p_{{344}}
    -p_{{115}}
    -p_{{125}}
    -p_{{235}}
    \right]\nonumber\\
    &+
    p_{{431}} 
    +p_{{521}}
    -p_{{622}}
    +p_{{413}}
    +p_{{513}}
    +p_{{643}}
    -p_{{524}} \nonumber \\ 
   & -p_{{634}}
    +p_{{245}} 
    +p_{{635}}
    -p_{{516}}
    -p_{{526}}
    +p_{{546}}
    \nonumber\\\nonumber
    &+5\left[
    p_{{442}}
    -p_{{324}}
    \right] \geq -6,
    \label{ineq:TNCQStab}
\end{align}
where states and effects are numbered according to the ordered set $\{\ketbra{+}{+},\ketbra{-}{-},\ketbra{+y}{+y},\ketbra{-y}{-y},\ketbra{0}{0},\ketbra{1}{1}\}$ and transformations are numbered according to the ordered set $\{I,\mathcal{Z},S,S^{-1}\}$. This inequality is violated by the stabilizer fragment above, for which ${\rm \beta}=-12$. 

This provides the first noise-robust  noncontextuality inequality for transformations that can be violated within in the single-qubit stabilizer theory.

Since quantum theory does not play any role in the derivation of this inequality, this constitutes a noise-robust nonclassicality witness for an arbitrary GPT fragment involving six GPT states, six GPT effects, and four GPT transformations satisfying the operational identities in Eq.~\eqref{eq:OperIdTransfStabilizer}.

\section{Applying our arguments beyond $\mathcal{PTM}$ scenarios}

Although our program is formulated for $\mathcal{PTM}$ scenarios, it is fairly easy to apply it to get some nonclassicality witnesses in more general circuits.
This is possible by lumping procedures in the circuit together to recover a $\mathcal{PTM}$ scenario. For instance, imagine one is interested in a $\mathcal{PT}_1\mathcal{T}_2\mathcal{M}$ scenario (where a preparation and measurement are implemented with two transformations in between, with the first drawn from a set $\mathcal{T}_1$ and the second from a set $\mathcal{T}_2$). One can then think of this as a $\mathcal{PT'M}$ where $\mathcal{T'}$ are transformations of the form $T_{1}\circ T_{2}$ where $T_{1}\in\mathcal{T}_1$ and $T_2\in\mathcal{T}_2$, as shown in Diagram~\eqref{diagram:  lumpingProcedure}. Having recovered a $\mathcal{PTM}$ scenario, one can then implement the linear programs described herein to find transformation noncontextuality inequalities. 

\begin{figure}[htb!]
\includegraphics[width=0.36\textwidth]{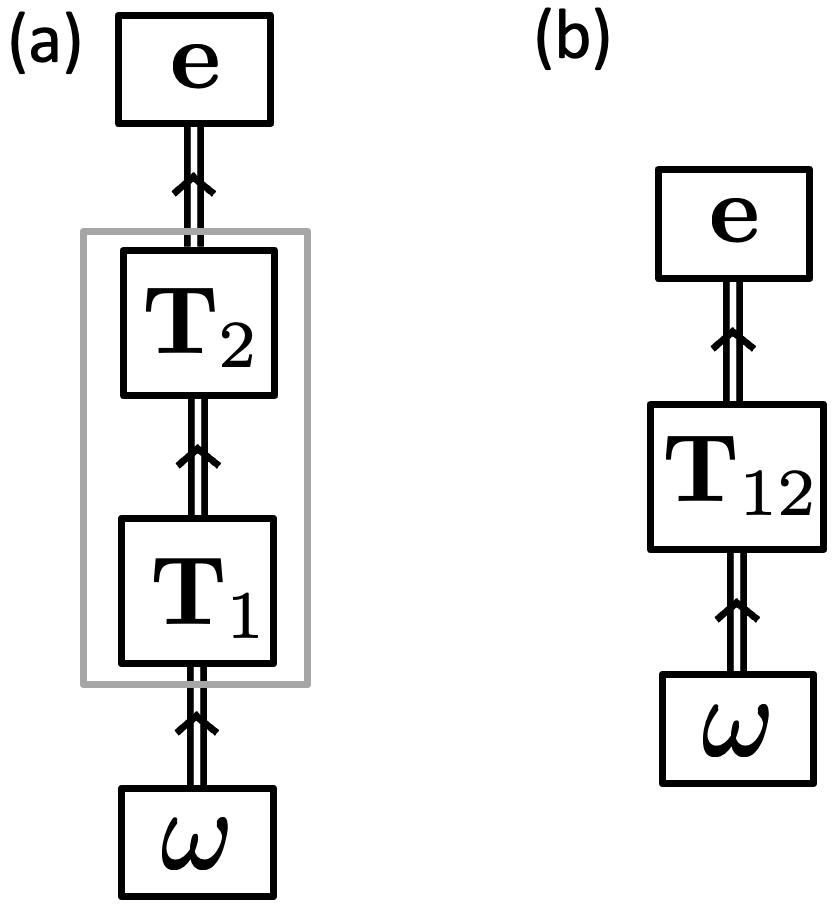}
\caption{(a) A  $\mathcal{PT}_1\mathcal{T}_2\mathcal{M}$  scenario. (b) Obtaining a $\mathcal{PTM}$ scenario by lumping the two transformations ${\bf T}_1$ and ${\bf T}_2$ to define ${\bf T}_{12}:={\bf T}_1\circ {\bf T}_2$.}
\label{diagram: lumpingProcedure}
\end{figure}

For instance, consider a $\mathcal{PT}_1\mathcal{T}_2\mathcal{M}$ fragment in the single qubit stabilizer theory, with the same states and measurements as in Section~\ref{sec:stabqubit}, but with the smaller set of transformations given by $\mathcal{T}_1=\{I,\mathcal{Z}\}$ and $\mathcal{T}_2=\{I,S\}$. In a $\mathcal{PTM}$ scenario where $\mathcal{T}=\{I,\mathcal{Z},S\}$, there are no nontrivial linear operational identities, so one might expect that our results cannot be nontrivially applied. However, using the lumping procedure just mentioned, we can define the set $\mathcal{T}'=\{I,\mathcal{Z},S,\mathcal{Z}\circ S\}$.  Since $\mathcal{Z}\circ S = S^{-1}$, we have $\mathcal{T}'=\{I,\mathcal{Z},S,S^{-1}\}$.  These four transformations satisfy a nontrivial operational identity (Eq.~\eqref{eq:OperIdTransfStabilizer}) and this identity allows for the derivation of a noncontextuality inequality that can be quantumly violated, as described in Sec~\ref{sec:stabqubit}.
This shows that the lumping procedure allows our program to find necessary conditions for noncontextuality, even for a $\mathcal{PT}_1\mathcal{T}_2\mathcal{M}$ fragment.

Note that this procedure does not leverage a {\em nonlinear} operational identity that is satisfied in this scenario, namely, $\mathcal{Z}\circ S = S^{-1}$.  That is, our program does not demand that the stochastic representation of $S^{-1}$ be equal to the stochastic representation of $\mathcal{Z}$ composed with that of $S$. 

In general, there are many different ways in which processes in a scenario can be lumped to obtain a $\mathcal{PTM}$ fragment. Even in a $\mathcal{PT}_1\mathcal{T}_2\mathcal{M}$  fragment, there are different meaningful lumping procedures that one could perform, arriving at (possibly) different $\mathcal{PTM}$ or $\mathcal{PM}$ fragments, depending on whether one lumps the transformations together with each other, with preparations, or with effects.
In principle, each of the fragments arriving from these lumping procedures could lead to different necessary conditions, some stronger than others.

Lumping processes together generically loses information, and consequently makes it easier to find an ontological model (and so harder to find proofs of nonclassicality). For instance, the stabilizer subtheory for a single qubit (including single-qubit transformations) does not admit of any ontological model. But when one lumps all of the transformations together with either the states or the effects, one obtains the prepare-measure scenario with all and only the stabilizer states and stabilizer effects on a qubit, which is noncontextual. 
So if one finds inequalities for a scenario by lumping together some of the processes, one should expect these inequalities to be weaker constraints on the existence of a noncontextual ontological model, and correspondingly less likely to be violated by one's theory or data. 
Therefore, if the data violates any noncontextuality inequality in the lumped scenario, then the original scenario necessarily also exhibits a violation of noncontextuality. 

Another way in which one can generalize beyond the PTM scenario considered here is to consider a PTM scenario for a {\em multipartite} system.  As shown in Ref.~\cite{zhang2026} (see the final paragraph of Sec.~IIC and the discussion in Sec.~V), assessments of nonclassicality  are always made relative to a given circuit structure (representing the causal structure), which includes a specification of subsystem structure.  For instance, a 4-dimensional quantum system can be modelled as a pair of qubits or as a single monolithic 4-level system, and assessments of nonclassicality can depend on which view one takes.  
Taking into account subsystem structure introduces additional operational identities and therefore additional constraints on the ontological model under the assumption of noncontextuality.  Nonetheless, treating a multi-partite system as a single monolithic system can still lead to nontrivial noncontextuality inequalities and their satisfaction still constitutes a necessary condition for classicality. Thus, any violation of these inequalities still certifies nonclassicality of the multipartite PTM scenario.

The point being made about lumping subsystems together into one system  is analogous to the point made above about lumping  sequences of processes together into one process.   Lumping systems together generically loses information, and consequently makes it easier to find an ontological model (and so harder to find proofs of nonclassicality), but if the data violates any noncontextuality inequality in the lumped scenario, then the original scenario necessarily also exhibits a violation of noncontextuality.

\section{Conclusions}

We have presented the first systematic method for witnessing nonclassicality in scenarios with transformations: a linear program that gives necessary conditions (inequalities) for the existence of a noncontextual model. Violations of these inequalities witness nonclassicality, regardless of the correctness of quantum theory. We have also derived the first noncontextuality inequality involving transformations that can be violated within the single-qubit stabilizer subtheory. 

There are a number of natural extensions of our work. Most obviously, one would like some general tools for studying scenarios with sequences of transformations or subsystem structure, and where general operational identities beyond the linear ones we have considered. Most likely, either of these generalizations makes the problem significantly more complicated, and it will no longer be a linear program (or indeed anything particularly tractable). In particular, incorporating the full set of operational identities, including those arising from compositional structure, leads to nonlinear constraints on the ontological model. Consequently, the set of correlations admitting of a noncontextual explanation is not expected to form a polytope, and therefore is unlikely to admit a characterization in terms of a finite set of linear inequalities, in contrast to the situation in Bell scenarios~\cite{Fine,Pitowsky}. Still, the question of finding or approximating the noncontextual set of correlations in an arbitrary scenario in a systematic (and tractable) manner is crucially important. Only with such tools can we fully realize the potential of tests of generalized noncontextuality as a universal method for understanding nonclassicality in all its forms, and in particular, its role in quantum computation.

It would be interesting to investigate whether the inequalities derived here admit of an operational interpretation in terms of communication or information-processing tasks, analogous to the connection between noncontextuality inequalities and random access codes identified in Ref.~[59]. Establishing such a connection in the present setting is nontrivial, due to our currently limited understanding of the resourcefulness of contextuality in transformations, and we leave this as an important direction for future work.

It is worth mentioning that a simple extension of our work that we do expect to be tractable (and still a linear program) is to cases where the transformations have outcomes, i.e., scenarios with a single instrument between the preparation and measurement stages of the experiment. We leave this generalization for future work. 

Another interesting direction would be to characterize the maximum values that quantum theory can achieve for the linear functional defining a given $\mathcal{PTM}$ inequality, generalizing known techniques for the prepare-measure case~\cite{Chaturvedi2021characterising,TavakoliBounding2021}.

\section*{Acknowledgements}
DS thanks Matt Leifer, Anubhav Chaturvedi, Tom{\'a}\v{s} Gonda, TC Fraser, Elie Wolfe, Ravi Kunjwal, Piers Lillystone, and Shane Mansfield for interesting discussions on this topic. 
DS and JHS were supported by the National Science Centre, Poland (Opus project, Categorical
Foundations of the Non-Classicality of Nature, Project No.
2021/41/B/ST2/03149). RDB and ABS acknowledge support by the Digital Horizon Europe project FoQaCiA, Foundations of quantum computational advantage, GA No. 101070558, funded by the European Union, NSERC (Canada), and UKRI (U.K.). RWS, RDB and DS are supported by the Perimeter Institute for Theoretical Physics. Research at Perimeter Institute is supported in part by the Government of Canada through the Department of Innovation, Science and Economic Development and by the Province of Ontario through the Ministry of Colleges and Universities. 
 
\section*{Author contribution statement}
All authors contributed to the ideas and results of the manuscript. DS, RDB, and JHS handled most of the mathematical proofs and writing of the manuscript. DS handled the journal submission. No AI tools were used at any stage of this project.

\bibliographystyle{quantum}
%\nocite{apsrev41Control}
\bibliography{bib}

\appendix

\section{Characterizing all the linear operational identities among any set of states, effects, or transformations}
\label{appendix:AllOpIdentitiesTransf}

We now present a systematic method for deriving a complete characterization of all the linear operational identities that hold among any given set of states, set of effects, or set of transformations. 
(As discussed in Section~\ref{sec:opidentities},
consideration of compositionality, such as subsystem structure,
leads to new possibilities for operational identities beyond linear constraints, but such possibilities are not considered here.)  As in the main text, we assume that one already has a GPT characterization of one's processes. See, e.g., Ref.~\cite{janotta2013generalized} for a discussion of how to represent quantum states in this manner, or Refs.~\cite{mazurek2021experimentally,grabowecky2021experimentally} for details on how such a characterization can be derived directly for experimental procedures in a theory-independent way.

We begin with the case of states.
Imagine one is given a finite set $\{{\bf s}_i\}_i$ of GPT state vectors. An operational identity among GPT states is a constraint of the form
\begin{equation}
\sum_i \alpha_i \bf{s}_i  = 0,
\end{equation}
for some set of real numbers $\{\alpha_i\}_i$. 
In general, there will be an infinite number of such constraints, but we can nonetheless characterize them by stipulating a finite set. 

To do so, we first construct a matrix $S$ whose $i$-th column is taken to be the vector ${\bf s}_i$. If there are $N$ GPT states in the given set, each of dimension $d$, then $S$ has dimensions $d \times N$. The rank of this matrix---the dimension of the vector space spanned by its rows (or equivalently, by its columns) is some number $k$ such that $k\leq N$ and $k \leq d$. If $k=N$, then there are no operational identities, as all the columns of the matrix must be linearly independent. But if $k < N$, then there must be linear dependences among the rows, and these linear relations define operational identities on the preparations. Recall that the kernel of a matrix is the linear subspace of vectors mapped to the zero vector by the matrix.
So, for any vector 
$\boldsymbol{\alpha}$ in the kernel of $S$, the equation 
\begin{align}
S\boldsymbol{\alpha} = \mathbf{0}
\end{align}
defines an operational identity, since
\begin{align}
&\left[
  \begin{array}{cccc}
    \vertbar & \vertbar &        & \vertbar \\
  {\bf  s}_{1}    & {\bf s}_{2}    & \ldots & {\bf s}_{N}    \\
    \vertbar & \vertbar &        & \vertbar 
  \end{array}
\right]
\left[
\begin{array}{c}
\alpha_1 \\ \alpha_2 \\ \vdots \\ \alpha_N
\end{array}
\right] \nonumber \\
&=
\left[
  \begin{array}{c}
  \hspace{-5mm}   \vertbar  \\
    \alpha_1   {\bf  s}_{1}+ \alpha_2   {\bf  s}_{2} + \ldots + \alpha_N  {\bf  s}_{N}    \\
  \hspace{-5mm}  \vertbar
  \end{array}
\right] \nonumber \\
 &= \sum_i \alpha_i  {\bf  s}_{i} = {\bf 0}
\end{align}

Any linear combination of vectors in the kernel is also in the kernel, so there are generally an infinite number of operational identities. However, one can restrict attention to any basis of linearly independent vectors  
$\{\boldsymbol{\alpha}^{(a)}\}_a$  which span $\textsf{ker}(S)$, where $a \in \{ 1,2, ..., k\}$. 

We refer to the set of operational identities corresponding to a basis of the kernel as a {\em generating set}.
This is an apt name because every such set satisfies the following crucial property:
{\em Preparation noncontextuality is satisfied with respect to the {\em full set} of operational identities among states if and only if it is satisfied with respect to {\em a generating set} of operational identities among the states.} 
We now prove this.
\begin{proof}
If preparation noncontextuality fails with respect to any operational identity in the generating set, then it trivially fails with respect to the full set of operational identities. Thus, we need only show that if preparation noncontextuality is satisfied with respect to the generating operational identities, then it must be satisfied with respect to all operational identities. 

Imagine one has a basis  
$\{\boldsymbol{\alpha}^{(a)}\}_a$  for $\textsf{ker}(S)$, so that the generating set of operational identities is
\beq \label{opeq_p}
\Big\{ \sum_i [\boldsymbol{\alpha}^{(a)}]_i \mathbf{s}_i = 0\Big\}_a
\eeq
where $[\mathbf{v}]_i$ denotes the $i$-th component of vector $\bf{v}$. Assume that the ontological model respects the operational identities in the generating set, so that one has
\beq 
\sum_i [\mathbf{\alpha}^{(a)}]_i \mu_{\mathbf{s}_i}(\lambda) = 0
\label{ineq:PNC}
\eeq
for all $\lambda$ and $a$. Now, for any $\mathbf{v}\in\textsf{ker}(S)$ one can write $\mathbf{v}=\sum_a v_a \boldsymbol{\alpha}^{(a)}$ for some real coefficients $\{v_a\}_a$, and it follows that
\begin{align}
\sum_i[\mathbf{v}]_i\mu_{\mathbf{s}_i}(\lambda) &= \sum_i  \left[\sum_a v_a \boldsymbol{\alpha}^{(a)}\right]_i\mu_{\mathbf{s}_i}(\lambda)\\ &= \sum_a v_a \sum_i[ \boldsymbol{\alpha}^{(a)}]_i\mu_{\mathbf{s}_i}(\lambda) \\
&= \sum_a v_a \cdot 0 \\
&= 0.
\end{align}
This means that if the model obeys the  implications of preparation noncontextuality for
the generating set of operational identities, then it necessarily obeys the implications of preparation noncontextuality for  
{\em all} operational identities in $\mathsf{ker}(S)$.   Equivalently, the failure of preparation noncontextuality relative to {\em any} operational identity exhibited by the data implies a failure of preparation noncontextuality for at least one of the operational identities in the generating set. 
\end{proof}

In summary, the satisfaction of preparation noncontextuality relative to a generating set of operational identities is both necessary and sufficient for satisfaction of preparation noncontextuality relative to all operational identities.

One can find all the operational identities among a set $\{{\bf e}_i\}_i$ of effects (on a single system with no subsystem structure) in the same manner. One first constructs the matrix $E$ whose $i$-th column is the GPT vector ${\bf e}_i$.  Any vector $\bf{v}$ in the kernel of $E$ defines an operational identity through the equation $ E \mathbf{v} = \mathbf{0}$. Any basis of linearly independent vectors which spans $\text{ker}(E)$ corresponds to a generating set of operational identities, in the sense that satisfaction of MNC relative to such a generating set is both necessary and sufficient for satisfaction of MNC relative to all operational identities.

The procedure for finding the set of all linear operational identities for transformations is also analogous. First, one finds a representation of the GPT transformations as vectors. (There are many ways to vectorize a matrix; most obviously, one can simply stack the columns of the matrix on top of one another.) Then, one constructs a matrix with these vectors as its columns and finds a basis for the kernel of that matrix; this basis corresponds to a generating set of operational identities.

\section{Proof of Formulation~\ref{c2}}
\label{appendix:mapLPs}

In the main text, we derived two linear programs, differing only by a factor of $N$. The first of these finds noncontextuality inequalities for the flag-convexified scenario, and we have claimed that the second finds noncontextuality inequalities for the original $\mathcal{PTM}$ scenario. We now prove this by considering some simple and general properties of linear programs.

Consider a linear program testing for the existence of $\bs{x}\geq \bs{0}$ such that
\begin{align}
    \mathds{M}\bs{x}=\bs{b}.
        \label{eq:appendixMxequalb2}
\end{align}
Such a solution exists if and only if the vector $\bs{b}$ satisfies all of the inequalities
\begin{align}
    \bs{y}^{(h)}\cdot \bs{b}\geq 0,\,\,\forall\, h\in\mathcal{H},
    \label{eq: appendixIneqsLP2}
\end{align}
where $\cdot$ represents the usual dot product. The set of inequalities, represented by the set of vectors $\{\bs{y}^{(h)}\}_h$, can be obtained by applying linear quantifier elimination to Eq.~\eqref{eq:appendixMxequalb2} to eliminate $\bs{x}$.

Now, consider some invertible matrix $D$ with the proper dimensions so that $D\mathds{M}$ is defined. Then,
\begin{gather}
   \left[\exists \bs{x}\geq \bs{0}\text{ s.t. }  D\mathds{M}\bs{x}=D\bs{b}\right] \nonumber \\
   \big\Updownarrow \nonumber \\   
    \left[\exists \bs{x}\geq\bs{0}\text{ s.t. } \mathds{M}\bs{x}=\bs{b}\right],
\end{gather}
since these two systems of linear equations are equivalent (due to the invertibility of $D$). This means that
\begin{gather}
    \left[\exists \bs{x}\geq \bs{0}\text{ s.t. }  D\mathds{M}\bs{x}=D\bs{b}\right] \nonumber \\
   \big\Updownarrow \nonumber \\ 
   \left[\bs{y}^{(h)}\cdot \bs{b}\geq 0,\,\,\forall\, h\in\mathcal{H}\right].
\end{gather}
In other words, since the existence problem involving $\mathds{M}$ and $\bs{b}$ is equivalent to the existence problem involving $D\mathds{M}$ and $D\bs{b}$, the set of inequalities in Expression~\eqref{eq: appendixIneqsLP2} provide necessary and sufficient conditions on $\bs{b}$ for $D\mathds{M}\bs{x}=D\bs{b}$ to admit of a solution. 
Note, however, that the set of vectors output by implementing quantifier elimination 
on $D\mathds{M}\bs{x} =D\bs{b}$ would in general yield a different (though equivalent) set of inequalities.  
This set can be obtained just by rewriting the inequalities~\eqref{eq: appendixIneqsLP2}, to get the equivalent ones:
\begin{gather}\label{jjj}
    \left[\bs{y}^{(h)}\cdot\bs{b}\geq 0,\,\,\forall\,h\in\mathcal{H}\right]\nonumber \nonumber \\
   \big\Updownarrow \nonumber \\   
    \left[\left(\bs{y}^{(h)}D^{-1}\right)\cdot D\bs{b}\geq 0,\,\,\forall\,h\in\mathcal{H}\right],
\end{gather}
which proves that the problem 
\begin{align}
    D\mathds{M}\bs{x}=\bs{b'},
\end{align}
with $\bs{b'}=D\bs{b}$, has a solution if and only if the following inequalities are satisfied:
\begin{align}       
 \left(\bs{y}^{(h)}D^{-1}\right)\cdot \bs{b'}\geq 0,\,\,\forall\, h\in\mathcal{H},
    \label{ineqs:appendixNewLP2}
\end{align}
where the set of vectors $\{\bs{y}^{(h)}\}$ is the one obtained  from linear quantifier elimination applied to the existence problem $\mathds{M}\bs{x}=\bs{b}$.

Now we are in place to show that, if Eqs.~\eqref{mc0fc}-\eqref{mc3fc} define a linear program for the flag-convexified scenario, then Eqs.~\eqref{mc0}-\eqref{mc3} define a linear program for the original scenario. Moreover, we obtain that the noncontextual bounds are just related by a factor of $N$.

Consider the program $\mathds{M}\bs{x} = \bs{b}$ defined by Eqs.~\eqref{mc0fc}-\eqref{mc3fc}:  vector $\bs{b}$ contains terms that depend on the flag-convexified probabilities $p(k\bar{s}|t)$ and some constant terms. Defining vector $\bs{b'}$ to be the one obtained from $\bs{b}$ by replacing the flag-convexified probabilities with the original ones, $p(k|st)$, we get
\begin{align}
    \bs{b'} = D\bs{b},
\end{align}
where $D$ is a diagonal matrix such that $D_{ii}=N$ , for indexes $i$ such that $b_i$ is some flag-convexified probability, and $d_{ii}=1$ for all other $i$. Note that this matrix is invertible. 
Therefore, given that $\mathds{M}\bs{x}=\bs{b}$ defines a linear program using the flag-convexified probabilities, $D\mathds{M}=\bs{b'}$ defines a linear program using the original probabilities. This maps Eqs.~\eqref{mc0fc}-\eqref{mc3fc} to Eqs.~\eqref{mc0}-\eqref{mc3} (given the form of $D$, the only difference in those linear programs is between Eqs.~\eqref{mc3fc} and ~\eqref{mc3}).

Finally, Eq.~\ref{jjj} specifies how the inequalities of the flag-convexified and original scenarios relate to each other given the form of matrix $D$:
\begin{gather}
    \left[\sum_{i\in I_0}y^{(h)}_ib_i + \sum_{i\not\in I_0} y^{(h)}b_i \geq 0\right]\nonumber 
   \nonumber \\
   \big\Updownarrow \nonumber \\ 
    \left[\sum_{i\in I_0}y^{(h)}_ib'_i + \frac{1}{N}\sum_{i\not\in I_0} y^{(h)}b'_i\geq 0\right],
\end{gather}
where we define $I_0$ to be the set of indices for which $b_i$ is independent of the probabilities (note that, given that $D$ is diagonal, $I_0$ is the same in the flag-convexified or original scenarios). This notational convention implies that one can rewrite inequalities  $\bs{y}^{(h)}\cdot\bs{b}\geq 0$ to get inequalities in the usual form
\begin{align}       
    \gamma^{(h)}_0 +\sum_{k,s,t}\gamma^{(h)}_{k,s,t}p(ks|t) \geq 0,
    \label{ineq:gammaineqsflag}
\end{align}
where $\gamma^{(h)}_{k,s,t}$ equals $y_i$ for $i\not\in I_0$ and $\gamma^{(h)}_0$ is the sum of all terms $y_ib_i$ with $i\in I_0$. Applying the same rewriting to the original scenario inequalities we have
\begin{align}
   &\gamma^{(h)}_0 +\frac{1}{N}\sum_{k,s,t}\gamma^{(h)}_{k,s,t}p(k|st)\geq 0\nonumber\\&\implies N\gamma^{(h)}_0 +\sum_{k,s,t}\gamma^{(h)}_{k,s,t}p(k|st)\geq 0.
\end{align}
We therefore see that the coefficients $\gamma_{k,s,t}^{(h)}$ for the original scenario are the same as the coefficients $\gamma_{k,s,t}^{(h)}$ for the flag-convexified scenario, but the noncontextual bound gets multiplied by a factor of $N$.

\end{document}